\documentclass[a4paper,fleqn,usenatbib]{mnras}

\pdfoutput=1

\usepackage[T1]{fontenc}
\usepackage{ae,aecompl}

\usepackage{graphicx}	
\usepackage{amsmath}	
\usepackage{txfonts}    

\defcitealias{KA97}{KA97}
\defcitealias{HK13}{Paper 1}
\defcitealias{HK14}{Paper 2}

\title[Numerical modelling of radio galaxy lobes]{Numerical modelling of the lobes of radio galaxies in cluster environments - III. Powerful relativistic and nonrelativistic jets}

\author[W. English, M. J. Hardcastle and M. G. H. Krause]{
W. English$^1$\thanks{E-mail: w.english@herts.ac.uk}, 
M. J. Hardcastle$^1$ and 
M. G. H. Krause$^{2,3,4,5}$
\\
$^1$Centre for Astrophysics Research, School of Physics, Astronomy and Mathematics, University of Hertfordshire, College Lane, Hatfield, Hertfordshire, AL10 9AB, UK\\
$^2$Universit\"ats-Sternwarte M\"unchen, Ludwig-Maximilians-Universit\"at, Scheinerstr. 1, 81679 M\"unchen, Germany\\
$^3$Max-Planck-Institut f\"ur extraterrestrische Physik, Giessenbachstr.~1, 85741 Garching, Germany\\
$^4$Excellence Cluster Universe, Technische Universit\"at M\"unchen, Boltzmannstrasse 2, 85748 Garching, Germany\\
$^5$School of Physical Sciences, University of Tasmania, Private Bag 37, Hobart, Tas 7001, Australia}

\date{Accepted XXX. Received YYY; in original form ZZZ}

\pubyear{2016}

\begin{document}
\label{firstpage}
\pagerange{\pageref{firstpage}--\pageref{lastpage}}
\maketitle

\begin{abstract}
We present results from two suites of simulations of powerful radio galaxies in poor cluster environments, with a focus on the formation and evolution of the radio lobes. One suite of models uses relativistic hydrodynamics and the other relativistic magnetohydrodynamics; both are set up to cover a range of jet powers and velocities. The dynamics of the lobes are shown to be in good agreement with analytical models and with previous numerical models, confirming in the relativistic regime that the observed widths of radio lobes may be explained if they are driven by very light jets. The ratio of energy stored in the radio lobes to that put into the intracluster gas is seen to be the same regardless of jet power, jet velocity or simulation type, suggesting that we have a robust understanding of the work done on the ambient gas by this type of radio source. For the most powerful jets we at times find magnetic field amplification by up to a factor of two in energy, but mostly the magnetic energy in the lobes is consistent with the magnetic energy injected. We confirm our earlier result that for jets with a toroidally injected magnetic field, the field in the lobes is predominantly aligned with the jet axis once the lobes are well developed, and that this leads to radio flux anisotropies of up to a factor of about two for mature sources. We reproduce the relationship between 151-MHz luminosity and jet power determined analytically in the literature.
\end{abstract}




\section{Introduction}

Due to the vast time-scales over which radio-loud active galaxies evolve, having accurate models to compare with observations is crucial. \citet{LRS73} set out several features that models must address to be consistent with observed radio galaxies: the central source must be constantly supplying energy to the lobes, and to avoid great adiabatic losses this energy must be in the form of relativistic beams or jets. Early analytical models of radio galaxies \citep{S74,BR74} used the spent jet material to inflate a broad cocoon encasing the jets. This cocoon is overpressured with respect to the ambient intracluster medium (ICM) at early times, causing it to expand supersonically and form a shell of shocked ICM. From here the models proceed in two ways: those in which the lobes remain overpressured and continue to expand supersonically with the jet confined by the cocoon material \citep[hereafter KA97]{BC89,KA97}, and those in which the internal pressure of the radio lobes comes into balance at some stage with the pressure of the ICM, and the lobes expand at constant pressure with energy continuously supplied by the central active galactic nucleus (AGN). Observational evidence suggests that the pressure inside the lobes is comparable to the thermal pressure of the external medium \citep{HW00,H02,C04}.

While these analytical models were being developed, high resolution observations of radio sources \citep{MKN68,M71} enabled \citet{FR74} to create a morphological classification for radio sources where a distinction is made between the centre-brightened class 1 (FRI) and edge-brightened class 2 (FRII) based on the location of the brightest region compared to the total length of the observed structure. This divide between FRI and FRII radio sources is seen to occur at a threshold luminosity (10$^{26}$ W Hz$^{-1}$ at 178 MHz) and was later reproduced in analytical models \citepalias{KA97} where less powerful jets (Q \textless {} 10$^{37}$ W) result in an FRI structure and more powerful jets (Q \textgreater {} 10$^{37}$ W) lead to an FRII.

While analytical models are a useful tool for understanding the basic physics of powerful radio galaxies, they necessarily make a number of assumptions regarding the symmetry and geometry of the sources. Numerical modelling essentially frees us of these constraints; we are instead limited by the demand for computing power by high resolution numerical models. Therefore simulations are required to make assumptions to cut down the simulation time, such as using simplified environments, non-relativistic jet speeds and over-dense jets, and have often ignored important physics such as magnetic fields, radiative losses and non-thermal particles. As computing resources become more readily available more sophisticated models are becoming more viable, giving a greater insight into the processes present in the lobes of radio galaxies.

In the past, environments have been taken to have a constant density profile \citep{N82,KM88,L89}, which may be acceptable for the early growth of the lobes but is clearly not over the length scales the radio lobes are observed to reach. This was later improved upon by considering more realistic beta model environments \citep{RHB02,BA03,K05} or an environment derived from a simulation of a dynamically active cluster \citep{H06}. Studies of the emission line gas around powerful radio sources showed that asymmetries in the distribution of ionized gas were correlated with the structural asymmetry of the radio lobes \citep{P89,MBK91}, which suggests that environmental asymmetries play an important role in creating structural asymmetries in the radio lobes. Numerical models of jets propagating through inhomogeneous environments are found to create strongly asymmetric lobes \citep{J05,GKC09,GKK11}.

Another factor that contributes to these asymmetries is relativistic beaming, an effect that has been observed, either as one-sided jets or non-unity jet-counterjet brightness ratios, for enough powerful active galaxies to suggest that all jets from AGN are relativistic \citep{C77,B89,WA97,H03,MH09}. Since this one-sidedness is often seen to occur for large length scales the jets in FRIIs must stay relativistic for the entire length of the lobes \citep{L93,MB09}, though evidence suggests that the termination shock at the end of the jets is not moving at highly relativistic speeds \citep{S95,DT97,AL00}; modestly relativistic lobe advance speeds are still possible in some sources. It is likely that a combination of relativistic beaming and environment are responsible for the observed asymmetries \citep{B95}. To cover these observed relativistic jet speeds, a number of models of AGN jets have been calculated with relativistic hydrodynamics (RHD) \citep[e.g.][]{G97,KF98,R99,P14}. Extending this to relativistic magnetohydrodynamics (RMHD) many models have been created focusing on the central region and the formation of the jets \citep[e.g.][]{KSK99,MKU01,N05,MB09,TNM11,MTB12}. Previous RMHD models of the evolution of the jets themselves have made the simplest possible assumption of injecting material with a toroidal magnetic field into a uniform unmagnetized \citep{J88,P96,M10} or magnetized \citep{N98,K99,L05,P15}, but otherwise simple medium, instead of using a more realistic environment.

Since the majority of the radiation from radio loud AGN comes from synchrotron emission from the jets and lobes, the inclusion of magnetic fields in models is essential for creating realistic synthetic observations. While emission maps have been created from purely hydrodynamical simulations by assuming the energy density in magnetic fields is proportional to the particle energy density \citep[e.g.][]{S85,S10,HK13}, this assumption is rather poor since the magnetic field is more intermittent than the pressure, as seen by filamentary structure in synchrotron radiation maps \citep{HC05}. When magnetic fields are included in the simulations the Stokes I, Q, and U parameters can be calculated along a given line of sight to give better synthetic observations \citep{CNB89,MS90a,MS90b,HEKA11,HK14}.

In this paper we follow on from the work presented in \citet[hereafter Paper 1 and Paper 2]{HK13,HK14}. In \citetalias{HK13} two-dimensional, purely HD models of the evolution of radio galaxy lobes in poor cluster environments were calculated in order to study the effect of environmental and jet properties on the resulting radio lobes. \citetalias{HK14} extended this to three-dimensional MHD models to obtain simulated observations, and look at how the environment affects the observational properties and magnetic field configuration. Here we present results from two suites of nine 3D simulations, each, of bipolar supersonic relativistic jets being injected into a realistic cluster environment, performed in RHD and RMHD, and study the effect of jet power and velocity on the evolution of the lobes. Section 2 describes the simulation setup for the models, in section 3 we present the results for the two suites and in section 4 we discuss the extent to which we believe them. In section 5 we summarize our findings.

\section{Simulation Setup}

The modelling in this paper makes use of the freely available code \textsc{pluto}\footnote{http://plutocode.ph.unito.it}, version 4.0, as described by \citet{M07}. The RHD and RMHD physics modules are both used, with the \textsc{hllc} (RHD) and \textsc{hlld} (RMHD) approximate Riemann solvers and a second order dimensionally unsplit Runge-Kutta time stepping algorithm, with a Courant-Friedrichs-Lewy number of 0.3. For the RMHD models a divergence cleaning algorithm is used to enforce $\nabla\cdot \mathbf{B}=0$. A Taub-Matthews equation of state is used to describe the relativistic gas with an adiabatic exponent that varies with temperature, from 4/3 (relativistic plasma) at high temperatures to 5/3 (ideal gas) at low temperatures \citep{MPB05,MK07}. The adaptive mesh refinement (AMR) capability of \textsc{pluto} is not used.

To avoid any numerical errors encountered when using extremely large or small units, \textsc{pluto} runs using simulation units. Three fundamental units (length, density and velocity) can be defined, from which all other units can be derived. For the relativistic modules the simulation unit of velocity $v_0$ must be the speed of light, $c$. For consistency with \citetalias{HK14}, we chose simulation units for length $L_0$ and density $\rho_0$ to be 2.1 kpc and 3.011$\times$10$^{-23}$ kg m$^{-3}$, respectively. Using a mean molecular weight $\mu$ of 0.6 we get a unit number density $n_j$ of 3$\times$10$^4$ m$^{-3}$. The remaining simulation units are derived from these by \textsc{pluto}, giving a simulation unit of pressure ($p_0=\rho_0 v_0^2$) which is $2.7\times10^{-6}$ Pa. In order to use the cluster environments of \citetalias{HK14} we are required to scale down all pressures by a factor of $2.7\times10^4$ so that the pressure at the centre of the cluster is 1 $p_0$, where $p_0 = 10^{-10}$ Pa. The simulation units for time ($t_0=L_0/v_0$) and magnetic field strength ($B_0=v_0\sqrt{4\pi\rho_0}$) are 6.85 kyrs and 1.84 $\mu$T.

The simulations model a 400 by 400 by 400 element volume ranging between $\pm$150 $L_0$, with periodic outer boundary conditions. This gives a physical resolution of ~1.6 kpc and allows the lobes to expand to a length of ~315 kpc. A cylindrical boundary condition with radius $r_j=2L_0$ and length $l_j=3L_0$ is defined at the centre of this volume aligned with the x-axis, and from this region the jet material is injected as an internal boundary condition. This results in a jet resolution of 2.7 cells per jet radius. Though our choice of value for $r_j$ is unphysically large, we are limited by the resolution of the simulations; the jet resolution has to be high enough that the internal boundary couples reasonably well with the environment. Internal to this boundary region the material is defined to have density $\rho_0$, velocity $v_j=v_x$ ($v_y=v_z=0$), temperature $T_j$, pressure $p_j=T_j\rho_j$ and, for the RMHD models, a purely toroidal magnetic field with $\mathbf{|\mathbf{B}|=B_j}$, where $\mathbf{B_y=B_j(z/r)}$ and $\mathbf{B_z=B_j(y/r)}$ for $\mathbf{r<r_j}$ (while a helical field structure might be more realistic, a toroidal field was used for simplicity and for consistency with the work described in \citetalias{HK14}). A conserved tracer quantity is also injected, initially with a value of 1.0 in the injection region and 0 everywhere else. The RHD models make use of \textsc{pluto}'s four-velocity module, which is unfortunately not available for the RMHD models.

The jet environment is that of a rich group or cluster, represented by an isothermal beta model, with density profile:

\begin{equation} \label{eq:betamod}
n=n_0\left[1+\left(\frac{r}{r_c}\right)^2\right]^{-\frac{3\beta}{2}}
\end{equation}
	
\noindent where in this paper the core radius $r_c$ is set to 30$L_0$, and $\beta$ has a value of 0.75. Small random perturbations are introduced to the density to break symmetry between the two lobes, though the same initial cluster environment is used for all of the runs so any differences arise naturally as a result of the different jet parameters. A vector gravitational force is defined by:

\begin{equation} \label{eq:gravpot}
\mathbf{g}=-\frac{3\beta}{\Gamma\times2.7\times10^4}\frac{1}{\sqrt{\mathbf{r}^2+r_c^2}}
\end{equation}

\noindent in order to keep the cluster environment stable, where the factor of $2.7\times10^4$ is the scaling factor applied to pressures throughout the models as described earlier in this section. A test simulation was carried out without the jet injection region and the cluster was seen to be stable for over 100,000 $t_0$ (0.67 Gyr), well beyond the duration of our simulations. The magnetic field in the cluster environment is set as a Gaussian random field, with an energy density that scales with thermal pressure, as described by \citet{M04} and \citet{H13}. Specifically, we generate the three components of the Fourier transform of the magnetic vector potential $A(k)$ by drawing their complex phases from a uniform distribution and their magnitudes from a Rayleigh distribution. The Fourier transform of the magnetic field is easily calculated from this, and by taking the inverse Fourier transform, and scaling to physical units, we are left with a divergence-free magnetic field with a peak strength at the centre of the cluster of $0.7$ nT.

In order to test the effects of jet power and velocity on the dynamics and energetics of the lobes and shocked gas a set of 9 simulations was created, covering 3 jet powers and 3 jet velocities. The jet power $Q$ for the two physics modules can be calculated, in SI units, with the following equations:

\begin{equation} \label{eq:RHDpow}
Q_{\text{RHD}}=\pi r_j^2v_j\left[\gamma(\gamma-1)\rho_jc^2+\frac{\Gamma}{\Gamma-1}\gamma^2P_j\right]
\end{equation}

\begin{align} \label{eq:RMHDpow}
Q_{\text{RMHD}}&=\pi r_j^2v_j\left[\gamma(\gamma-1)\rho_jc^2+\frac{\Gamma}{\Gamma-1}\gamma^2P_j+v_j\gamma_j^2\frac{B_j^2}{\mu_0}\right]
\end{align}

\noindent where $\gamma$ is the Lorentz factor and $\Gamma$ is the adiabatic index. For the RMHD model since we are injecting material with $v_y=v_z=0$ and a toroidal magnetic field around the x-axis, $\mathbf{v}\cdot\mathbf{B}=0$ and so this term is only included in equation 4 for completeness. To keep the jet power constant as the velocity is varied, two variables can be changed: $\rho_j$ and $T_j$ (though with higher resolution models we could also vary $r_j$). To remove some of the variability we set the ratio of the contributions to jet power from kinetic energy and enthalpy to unity, so that for a given jet power and velocity there is a single pair of values for $\rho_j$ and $T_j$. Unfortunately this limits our study to more powerful jets than discussed in papers 1 and 2, since we found that the lighter, faster jets with powers lower than $\sim1\times10^{39}$W could not overcome the ram pressure at the centre of the cluster, and therefore would not expand out of the injection region. With higher resolution models we could lower the jet power by decreasing the radius of the injection region, while keeping the velocity, density and temperature of the jet at the values used here. With the higher jet velocities afforded us by the relativistic modules we can model jets with more realistic densities for a given jet power, with the jets being under-dense when compared to their environment by up to a factor of $10^5$ \citep[eg.][]{K03, K05}. The values used for the RHD and RMHD models are presented in Tables~\ref{tab:RHD} and ~\ref{tab:RMHD}, respectively, where $M$ is the internal mach number and $\eta_r$ is the relativistic generalisation of the density contrast between the jet and the ambient medium $\eta$, and is given by:

\begin{equation}
\eta_r=\frac{\rho_jh\gamma^2}{\rho_a} 
\end{equation}

\noindent where $\rho_a$ is the ambient density and $h$ is the specific enthalpy \citep{K05}. Since the density at the environment at the centre of the cluster is 1, in simulation units, the densities presented here also show the values of $\eta$.

\begin{table*}
\begin{center}
\caption{RHD Simulation jet paramters. From left to right the columns give the code used to identify each model, the jet power $Q$, the injection velocity $v_j$, the Lorentz factor $\gamma_j$, the Mach number $M_j$, the jet density $\rho_j$, the ratio of jet and ambient pressure at the cluster centre, relativistic density contrast $\eta_r$, and the temperature $T_j$.}
\begin{tabular}{ ccccccccc }
\hline
Code & $Q$ & $v_j$ & $\gamma_j$ & $M_j$ & $\rho_j$ & $P_j/P_a$ & $\eta_r$ & $T_j$\\
& (W) & ($c$) & & & (Sim. units) & & (Sim. units) & (Sim. units)\\
\hline
v25-low & $1\times10^{39}$ & $0.25$ & 1.03 & 0.75 & $1.469\times10^{-3}$ & 0.40 & $2.192\times10^{-3}$ & $2.691\times10^2$\\
v60-low & $1\times10^{39}$ & $0.60$ & 1.25 & 1.80 & $1.002\times10^{-4}$ & 0.13 & $2.641\times10^{-4}$ & $1.284\times10^3$\\
v95-low & $1\times10^{39}$ & $0.95$ & 3.20 & 2.85 & $2.345\times10^{-5}$ & 0.06 & $4.605\times10^{-4}$ & $2.478\times10^3$\\
v25-med & $2\times10^{39}$ & $0.25$ & 1.03 & 0.75 & $2.939\times10^{-3}$ & 0.79 & $4.385\times10^{-3}$ & $2.691\times10^2$\\
v60-med & $2\times10^{39}$ & $0.60$ & 1.25 & 1.80 & $2.005\times10^{-4}$ & 0.26 & $5.283\times10^{-4}$ & $1.284\times10^3$\\
v95-med & $2\times10^{39}$ & $0.95$ & 3.20 & 2.85 & $4.690\times10^{-5}$ & 0.12 & $9.211\times10^{-4}$ & $2.478\times10^3$\\
v25-high & $5\times10^{39}$ & $0.25$ & 1.03 & 0.75 & $7.348\times10^{-3}$ & 1.98 & $1.096\times10^{-2}$ & $2.691\times10^2$\\
v60-high & $5\times10^{39}$ & $0.60$ & 1.25 & 1.80 & $5.011\times10^{-4}$ & 0.64 & $1.321\times10^{-3}$ & $1.284\times10^3$\\
v95-high & $5\times10^{39}$ & $0.95$ & 3.20 & 2.85 & $1.173\times10^{-4}$ & 0.29 & $2.304\times10^{-3}$ & $2.478\times10^3$\\
\hline
\end{tabular}
\label{tab:RHD}
\end{center}
\end{table*}

\begin{table*}
\begin{center}
\caption{RMHD Simulation paramters. Columns are the same as in Table~\ref{tab:RHD}, along with the strength of the injected magnetic field $B_j$ and plasma $\beta=2\times P_j/B_j^2$}.
\begin{tabular}{ ccccccccccc }
\hline
Code & $Q$ & $v_j$ & $\gamma_j$ & $M$ & $\rho_j$ & $P_j/P_a$ & $\eta_r$ & $T_j$ & $B_j$ & $\beta$\\
& (W) & ($c$) & & & (Sim. units) & & (Sim. units) & (Sim. units) & (Sim. units) & \\
\hline
v25-low-m & $1\times10^{39}$ & $0.25$ & 1.03 & 0.75 & $1.376\times10^{-3}$ & 0.43 & $2.053\times10^{-3}$ & $3.122\times10^2$ & $2.66\times10^{-4}$ & 107.13\\
v60-low-m & $1\times10^{39}$ & $0.60$ & 1.25 & 1.80 & $6.216\times10^{-5}$ & 0.12 & $1.638\times10^{-4}$ & $1.966\times10^3$ & $2.20\times10^{-4}$ & 44.55\\
v95-low-m & $1\times10^{39}$ & $0.95$ & 3.20 & 2.85 & $1.739\times10^{-6}$ & 0.01 & $3.416\times10^{-5}$ & $6.762\times10^3$ & $8.59\times10^{-5}$ & 28.12\\
v25-med-m & $2\times10^{39}$ & $0.25$ & 1.03 & 0.75 & $2.752\times10^{-3}$ & 0.86 & $4.106\times10^{-3}$ & $3.122\times10^2$ & $3.64\times10^{-4}$ & 114.42\\
v60-med-m & $2\times10^{39}$ & $0.60$ & 1.25 & 1.80 & $1.243\times10^{-4}$ & 0.24 & $3.252\times10^{-4}$ & $1.966\times10^3$ & $2.91\times10^{-4}$ & 50.55\\
v95-med-m & $2\times10^{39}$ & $0.95$ & 3.20 & 2.85 & $3.478\times10^{-6}$ & 0.02 & $6.831\times10^{-5}$ & $6.762\times10^3$ & $1.14\times10^{-4}$ & 31.93\\
v25-high-m & $5\times10^{39}$ & $0.25$ & 1.03 & 0.75 & $6.882\times10^{-3}$ & 2.15 & $1.026\times10^{-2}$ & $3.122\times10^2$ & $5.75\times10^{-4}$ & 114.67\\
v60-high-m & $5\times10^{39}$ & $0.60$ & 1.25 & 1.80 & $3.108\times10^{-4}$ & 0.61 & $8.190\times10^{-4}$ & $1.966\times10^3$ & $4.60\times10^{-4}$ & 50.95\\
v95-high-m & $5\times10^{39}$ & $0.95$ & 3.20 & 2.85 & $8.696\times10^{-6}$ & 0.06 & $1.328\times10^{-4}$ & $6.762\times10^3$ & $1.79\times10^{-4}$ & 32.38\\
\hline
\end{tabular}
\label{tab:RMHD}
\end{center}
\end{table*}

\begin{figure}
\includegraphics[width=84mm]{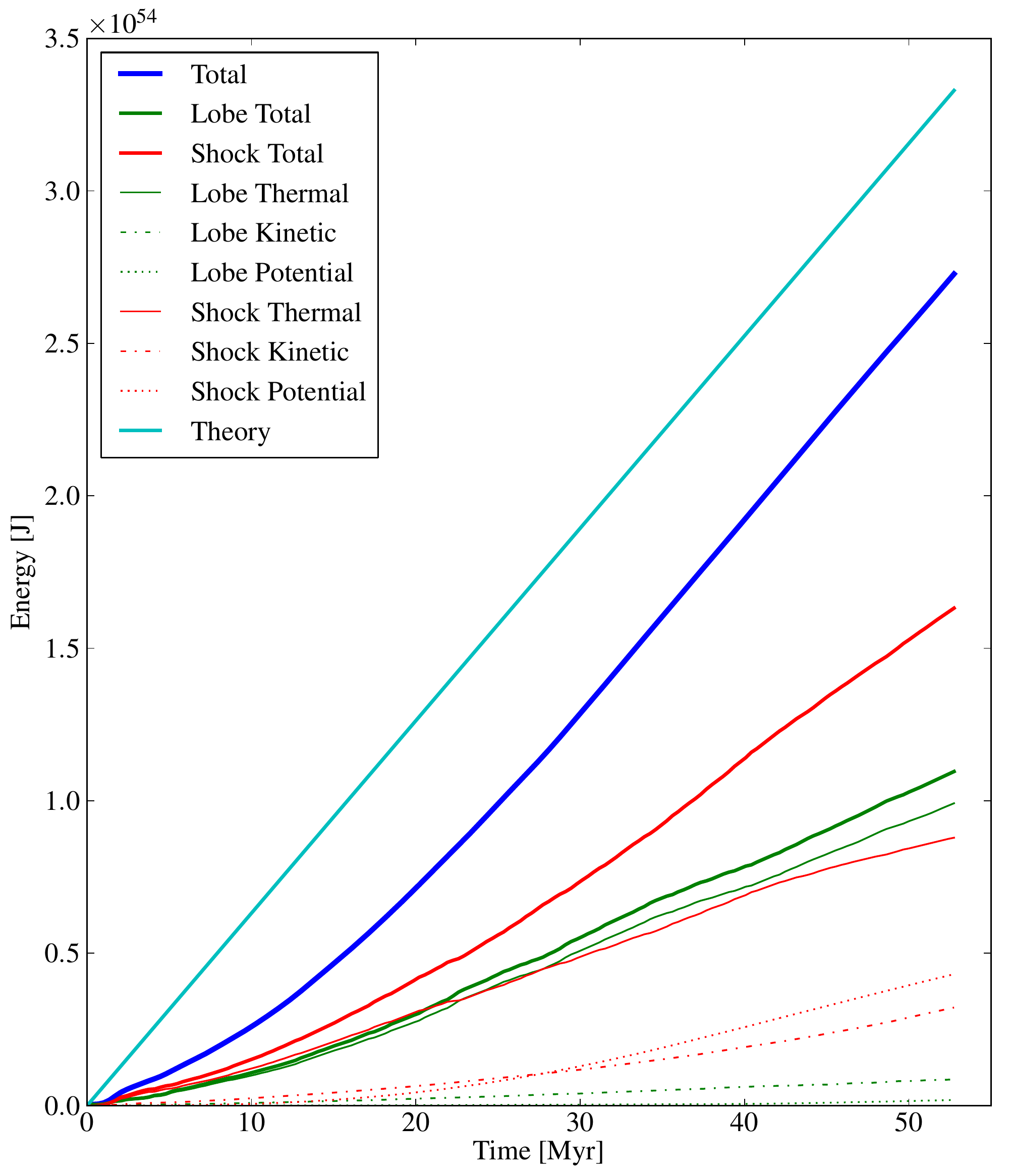}
\caption{Energies stored in the shocked and lobe regions for the \textbf{v60-med} simulation. The expected total energy is also plotted, which assumes the amount of energy flowing into the system is constant and is given by equation 3. This model is seen to be well coupled with the environment by 8.3 Myr.}
\label{fig:power}
\end{figure}

One potential issue comes from how \textsc{pluto} handles material falling into the side of the cylindrical injection region. The properties of the material here are set at each time-step to the jet values, so any material entering this region will vanish. At early times, back-flowing material from the two lobes merges close to the injection region resulting in some material being driven vertically outwards and some material vanishing into the injection region. The material pushed outwards goes on to form a structure between the two lobes which, while unphysical, is included in the lobe region when calculating the energetics of the system as the tracer quantity is greater than our threshold. This should have little effect on our results due to the relative volumes of the regions. The material that vanishes into the injection region carries some amount of energy with it, resulting in a discrepancy between the energy injected and the energy accounted for in the models. In addition to this, poor coupling between the internal boundary and the external conditions results in suppression of the flow from the injection region until the jet is well established. These effects are most notable at early times, as seen in Fig.~\ref{fig:power}, where the total amount of energy entering the system is systematically lower than the theoretical power. This is less of an issue later when the gradient of the total energy is comparable to the expected value from equations ~\ref{eq:RHDpow} and ~\ref{eq:RMHDpow}. To account for this the age of each model is counted from the time that the jet is well coupled with the cluster environment, calculated by extrapolating the total energy curve back (once it has stabilized).

All of the runs were carried out on the Science and Technology Research Institute (STRI) cluster of the University of Hertfordshire\footnote{http://stri-cluster.herts.ac.uk}. Each job was run on 192 Xeon-based cores, taking between 1-14 days each for the final runs (lower resolution runs were done, with fewer cores, in order to test the relativistic implementation and configure variables), with each of the RMHD models taking around 3 times as long as the corresponding RHD model. The Message Passing Interface (MPI) was implemented in \textsc{MVAPICH2}. An output file was written by \textsc{pluto} every 50 simulation time units, or every 0.34 Myr, consisting of density, velocity, pressure, magnetic field strength and tracer quantity values for the whole simulation grid. These values were used to compute the properties of the lobes, and the amount of energy stored in the lobes and shocked region. The region defined as the lobes is found by searching, from the outside of the grid inwards, for the surface where the tracer quantity has a value of 10$^{-3}$. The two lobes are processed independently, as the initial conditions on either side of the source are slightly different, though values for lobe properties stated will be an average value for the two lobes unless explicitly stated otherwise. The jet material is identified in the same way, but with the tracer threshold set to $0.9$. The shocked ambient gas is identified in a similar way; as the surface where the radial velocity is equal to $2.5\times10^{-4}$ $c$ (this threshold is found to be robust at identifying the shocked region, and is considerably larger than the velocity noise in the undisturbed environment). The initial thermal energy for all material within the shocked ambient gas region is subtracted from its thermal energy, to compensate for the latent thermal energy in the cluster material.

\section{Results}

For the remainder of this paper results will be discussed in physical units, using the conversion factors described in section 2. We will begin by discussing the results of the RHD simulations, followed by the results from the RMHD models and the synthetic observations of these models.

It is important to note here the effect that having poor resolution in the jet may have on our results. While we are confident this will not greatly affect the dynamics of the lobes, which is the main focus of this paper, this low resolution means that the jet is subject to strong numerical diffusion, which could prevent the onset of turbulence and instabilities in the jet which are key factors in amplifying the magnetic field. It is also worth noting that, while similar to the models of \citetalias{HK13} and \citetalias{HK14}, these sources are much more powerful (by up to a factor of 50) due to the limits described in the previous section, which should be taken into account when comparing these results, though they are comparable in terms of jet power to some of the models by \citet{HEKA11}.

\subsection{Relativistic Hydrodynamics}

\begin{figure*}
\includegraphics[width=175mm]{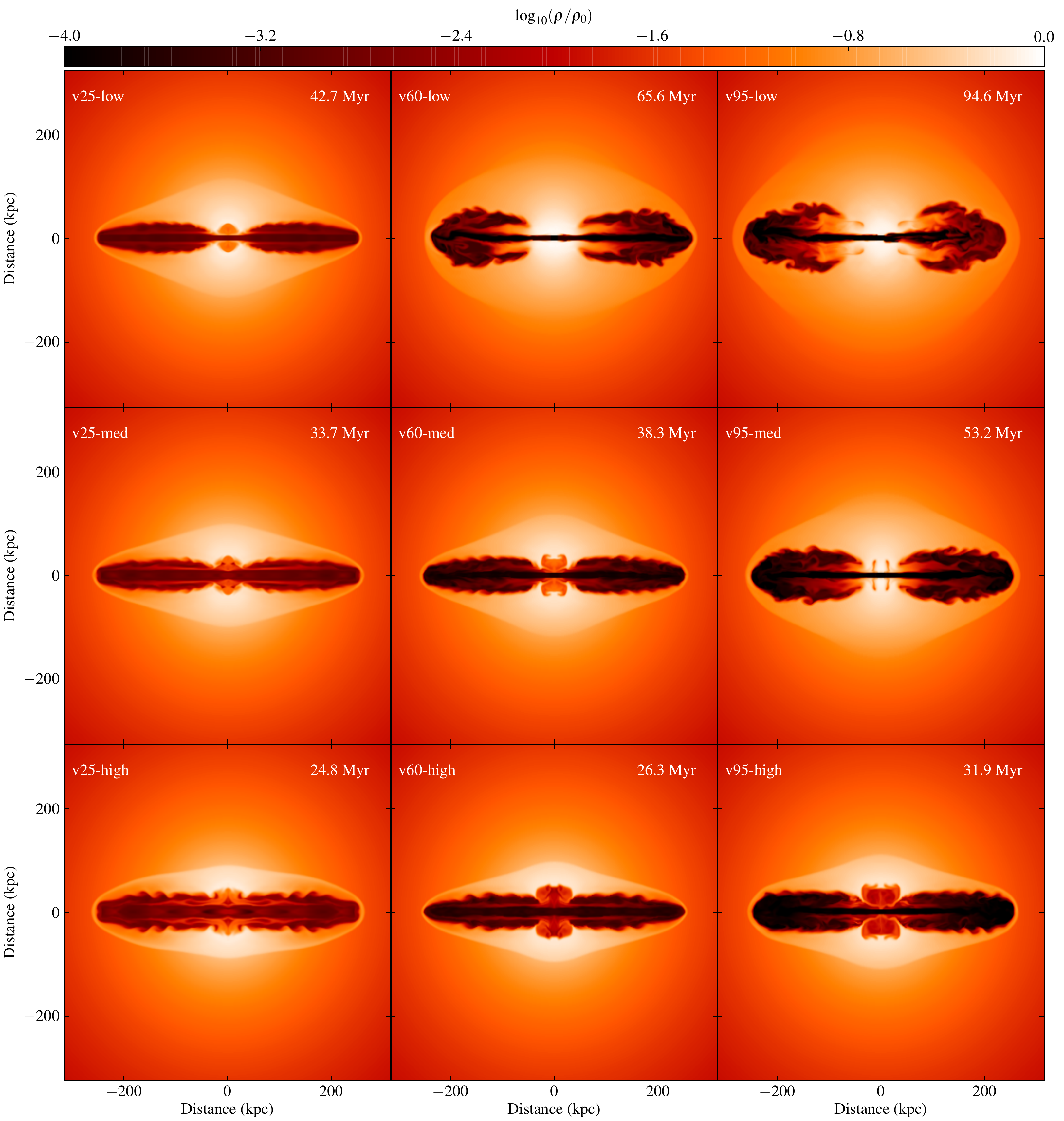}
\caption{Midplane density slices for the suite of RHD simulations, taken when the average length of the two lobes is 250 kpc. Top row: low power ($Q=1\times10^{39}$ W) models with jet velocities 0.25$c$, 0.6$c$ and 0.95$c$, respectively. Middle row: medium power ($Q=2\times10^{39}$ W) models with jet velocities 0.25$c$, 0.6$c$ and 0.95$c$, respectively. Bottom row: high power ($Q=5\times10^{39}$ W) models with jet velocities 0.25$c$, 0.6$c$ and 0.95$c$, respectively. Colour scale is logarithmic in simulation units of density, ranging from $-4$ (black) to $0$ (white). The label in the top left corner of each plot gives the age of each model at the time of the snapshots.}
\label{fig:RHDdens}
\end{figure*}

\begin{figure*}
\includegraphics[width=175mm]{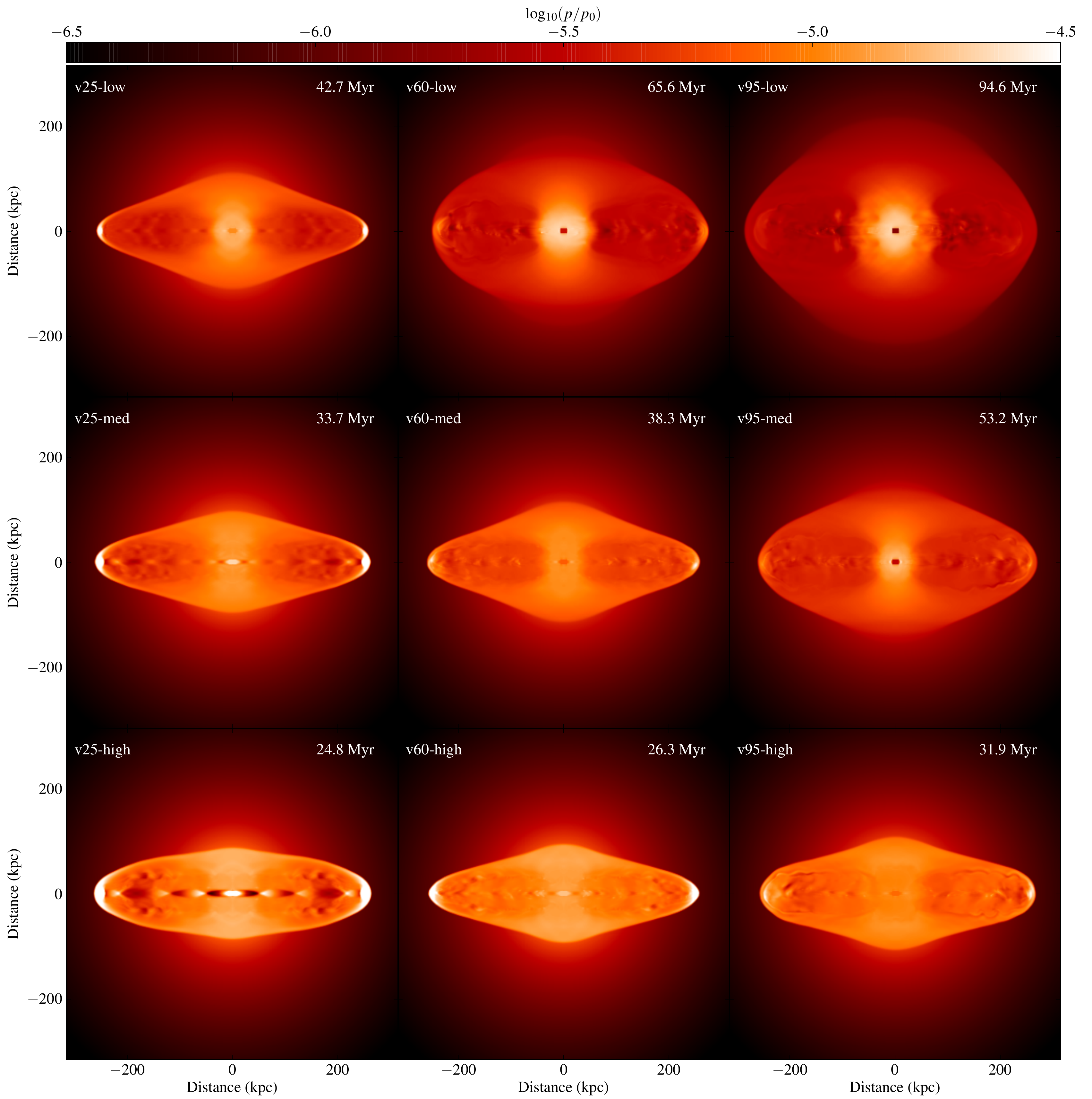}
\caption{Midplane pressure slices for the suite of RHD simulations, taken at the same time as Fig.~\ref{fig:RHDdens} and covering the same jet parameters. Colour scale is logarithmic in simulation units of pressure, ranging from $-6.5$ (black) to $-4.5$ (white).}
\label{fig:RHDprs}
\end{figure*}

Figures~\ref{fig:RHDdens} and~\ref{fig:RHDprs} show snapshot density and pressure maps as slices through the central $xy$ plane for all the simulations in the suite, for the time when the average length of the lobes is 250 kpc, showing the different morphologies that arise in the different runs. The lobes can be identified as the broad low density regions on either side of the central source, connected by the narrow jets. Surrounding this can be seen the strong shock propagating through the ambient medium. Any asymmetry in the lobes for a given source forms naturally due to the slight density perturbations in the initial environment, with the injection region boundary conditions being identical on either side of the source. The structure around the edges of the lobes are caused by Kelvin-Helmholtz instabilities and are generally unobserved in real radio galaxies. The pressure maps show some evidence for a jet termination shock forming, appearing as a high pressure surface at the ends of the jets just inside the main shocked region which is strongest for the slower, higher power jets.

\begin{figure}
\includegraphics[width=84mm]{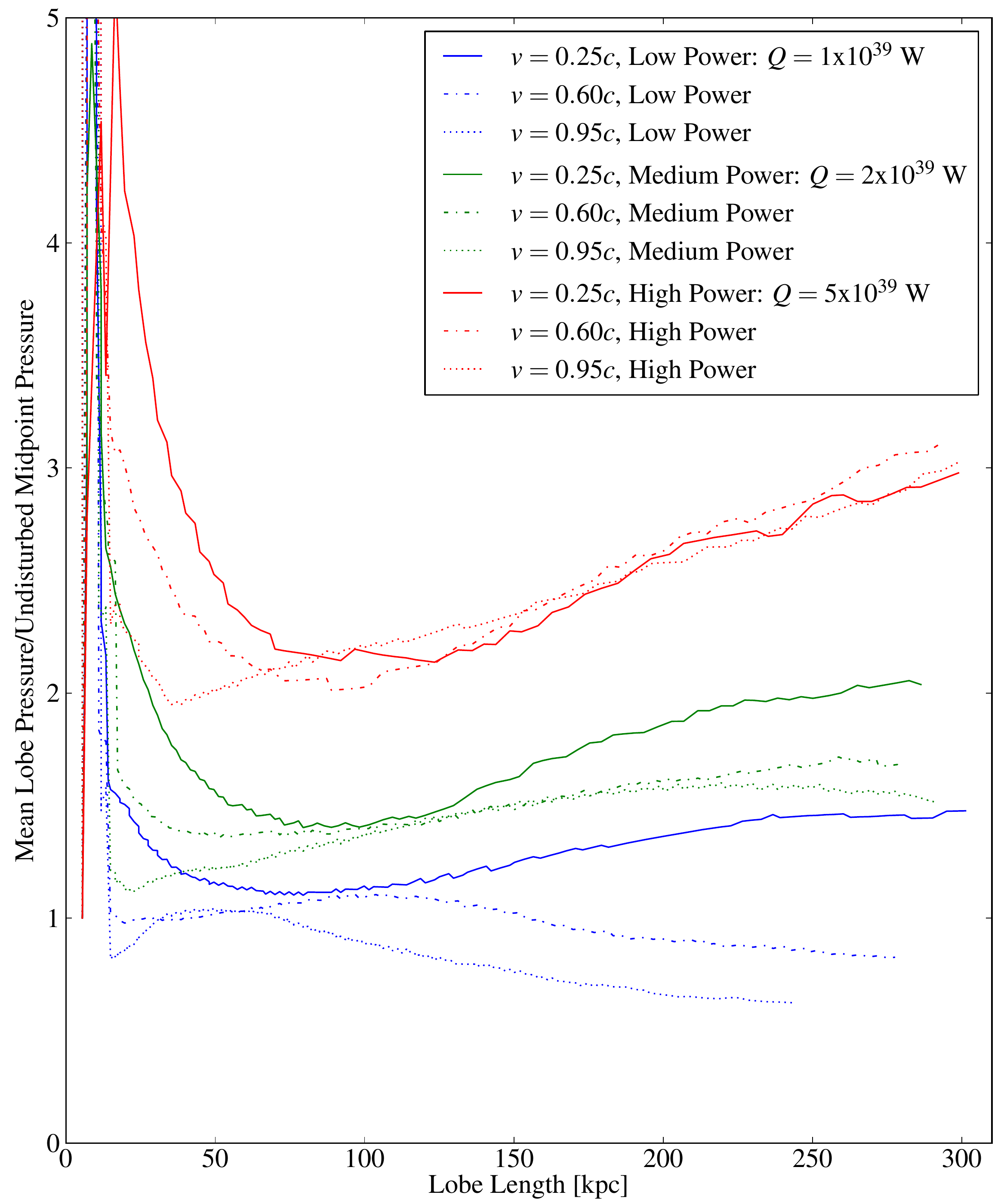}
\caption{The ratio between the mean pressure in the lobes and the \textit{undisturbed} external pressure at the midpoint of the lobes as a function of lobe length in kpc, for the RHD models.}
\label{fig:pressure}
\end{figure}

As expected, we see that the slower, denser jets lead to lobes that are less turbulent than for the relatively lighter jets, especially in the back-flowing jet material. The lobes for the low power runs are seen (Fig.~\ref{fig:pressure})  to stay in rough pressure equilibrium with the external medium (as meassured at the midpoint of the lobes) for most of the evolution of the source, whereas the higher power models are overpressured, by up to a factor of $3$, for the whole evolution, and become increasingly over-pressured as the lobes grow and leave the core of the cluster.

\begin{figure*}
\includegraphics[width=175mm]{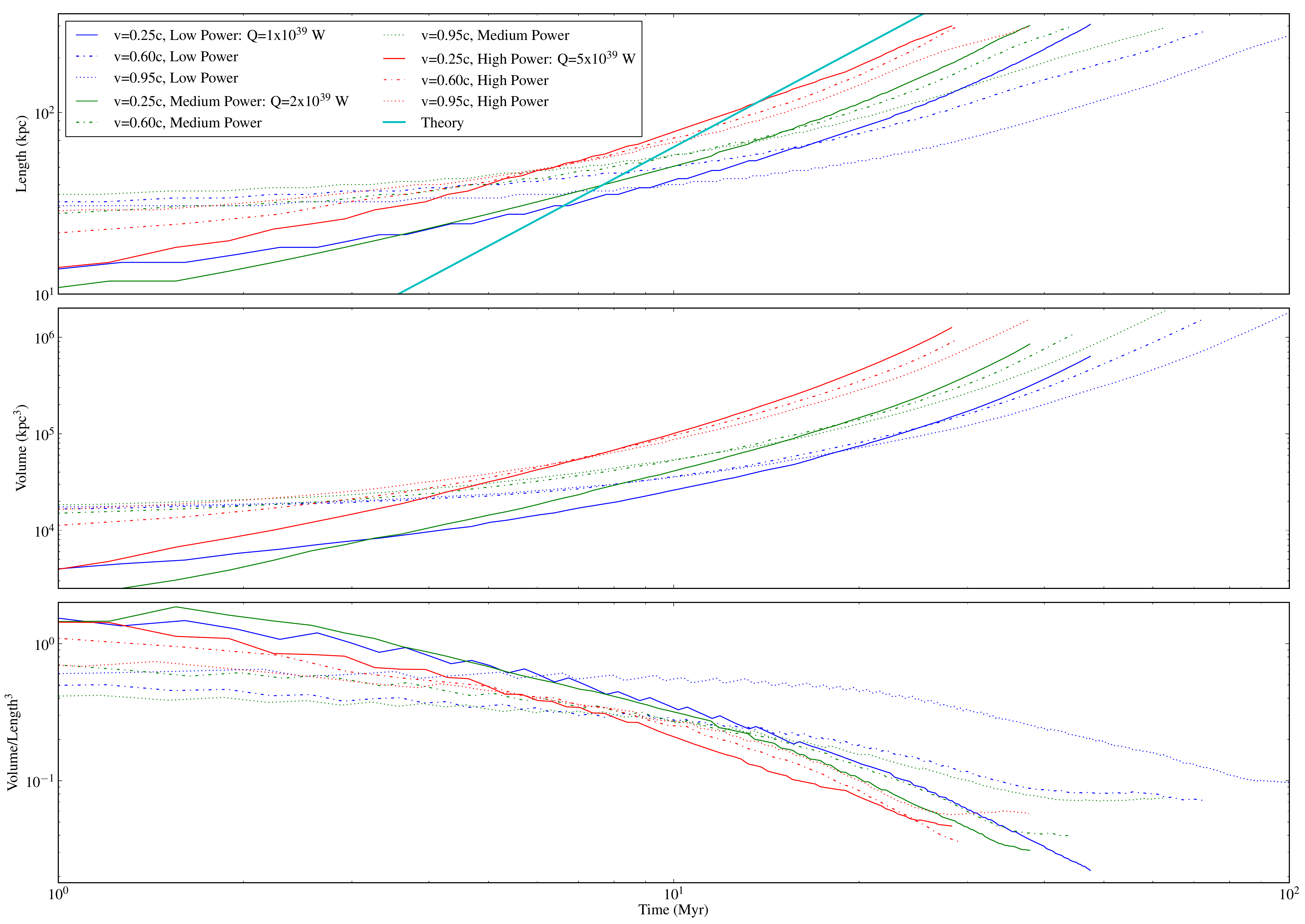}
\caption{Growth of the lobes with time for the RHD models. Time $t = 0$ is taken to be the time that the model is well coupled with the environment. For each model the plotted value is the average value for the two lobes. The theory line is the predicted growth of the lobe length from \citetalias{KA97}.}
\label{fig:RHDdyn}
\end{figure*}

For all of the simulations the length of the lobes is initially seen to have linear growth (Fig.~\ref{fig:RHDdyn}), before steepening at later times once leaving the central core to approach the predicted slope, for a ram pressure balanced jet pushing through a power law atmosphere, of $5/(5-3\beta)$ \citepalias{KA97}. As expected the slower jets lead to faster growing lobes, as in order to keep the power constant these are necessarily denser and therefore have a higher momentum flux for the same kinetic energy flux. Likewise the more powerful jets also lead to faster growth, and at higher powers the spread between the different speeds decreases.

The same relationship is seen in the growth of the volume of the lobes, with slower, more powerful jets having the fastest growth at early times, though later the fast light jets start to catch up and result in significantly wider lobes by the time the shock reaches the edge of the grid and the simulation ends.

The volume of the lobes grows with the length, but not self-similarly, as shown in the bottom panel of Fig.~\ref{fig:RHDdyn}. Here we see that while the lobes are initially quasi-spherical, the ratio of volume to length cubed starts to fall, as the jet forces the length of the lobes to grow faster than the lobe's lateral expansion. The faster jets, with slower lobe growth, are seen to have a higher ratio at all times, as they are more efficiently slowed by the ICM and have more spent jet material (jet material that has reached the end of the jet and been sufficiently slowed by the environment) inflating the lobes for a given lobe length. There is evidence from some of the simulations that this ratio flattens off at very late times.

\begin{figure*}
\includegraphics[width=175mm]{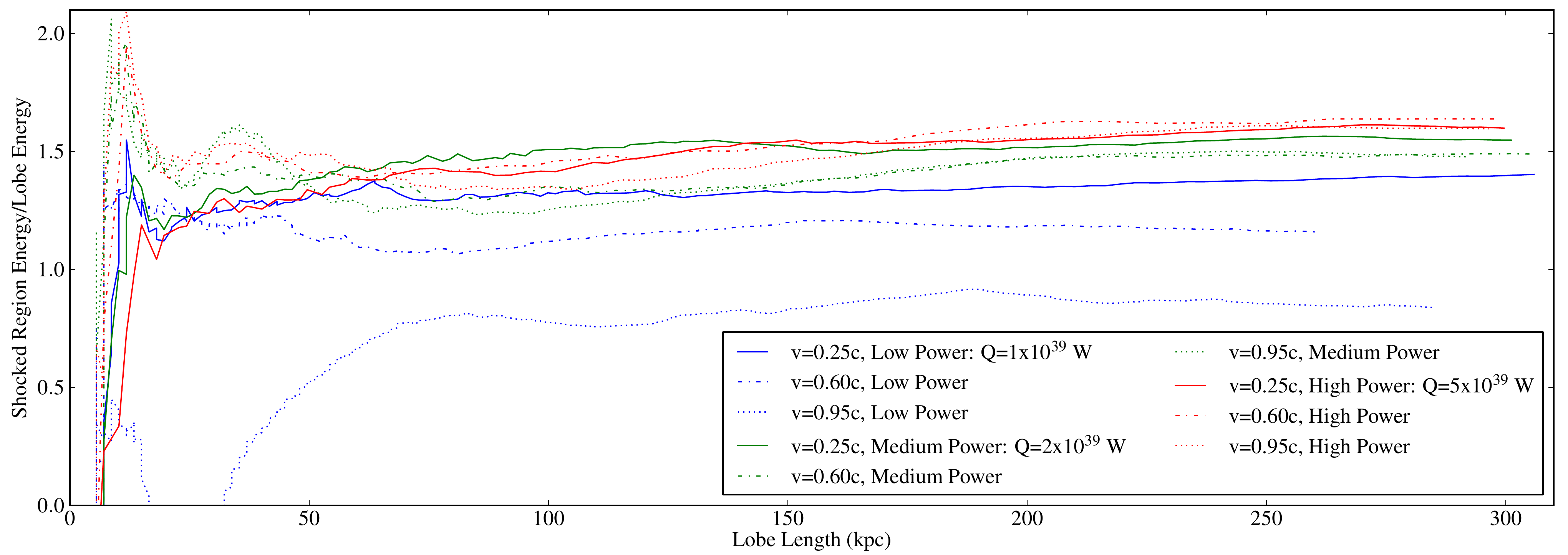}
\caption{Evolution of the ratio of energy stored in the shocked region to that in the lobes, as a function of the length of the lobes, for the RHD models.}
\label{fig:RHDenergy}
\end{figure*}

Looking at the ratio of energy stored in the lobes to that stored in the shocked material, shown in Fig.~\ref{fig:RHDenergy}, we see that the majority of the models have a ratio between 1.4 and 1.6 for almost the whole evolution of the simulations, with only the \textbf{v95-low} and \textbf{v60-low} simulations lying outside this range. All of the simulations are within the 0.6 to 1.8 range that was seen for the MHD models \citepalias{HK14}, and the RHD models show much less spread at late times than the MHD models.

\subsection{Relativistic Magnetohydrodynamics}

The RMHD models show the same general structure as the RHD ones though less uniform at all times, likely due to these models having slightly lighter jets which are more strongly affected by the perturbations in the environment. The magnetic field being injected has too low an energy density to be dynamically important with the energy in the magnetic fields being around $0.01$ times the thermal energy in the particles, though the field can still be locally dynamically non-negligible.

\begin{figure*}
\includegraphics[width=175mm]{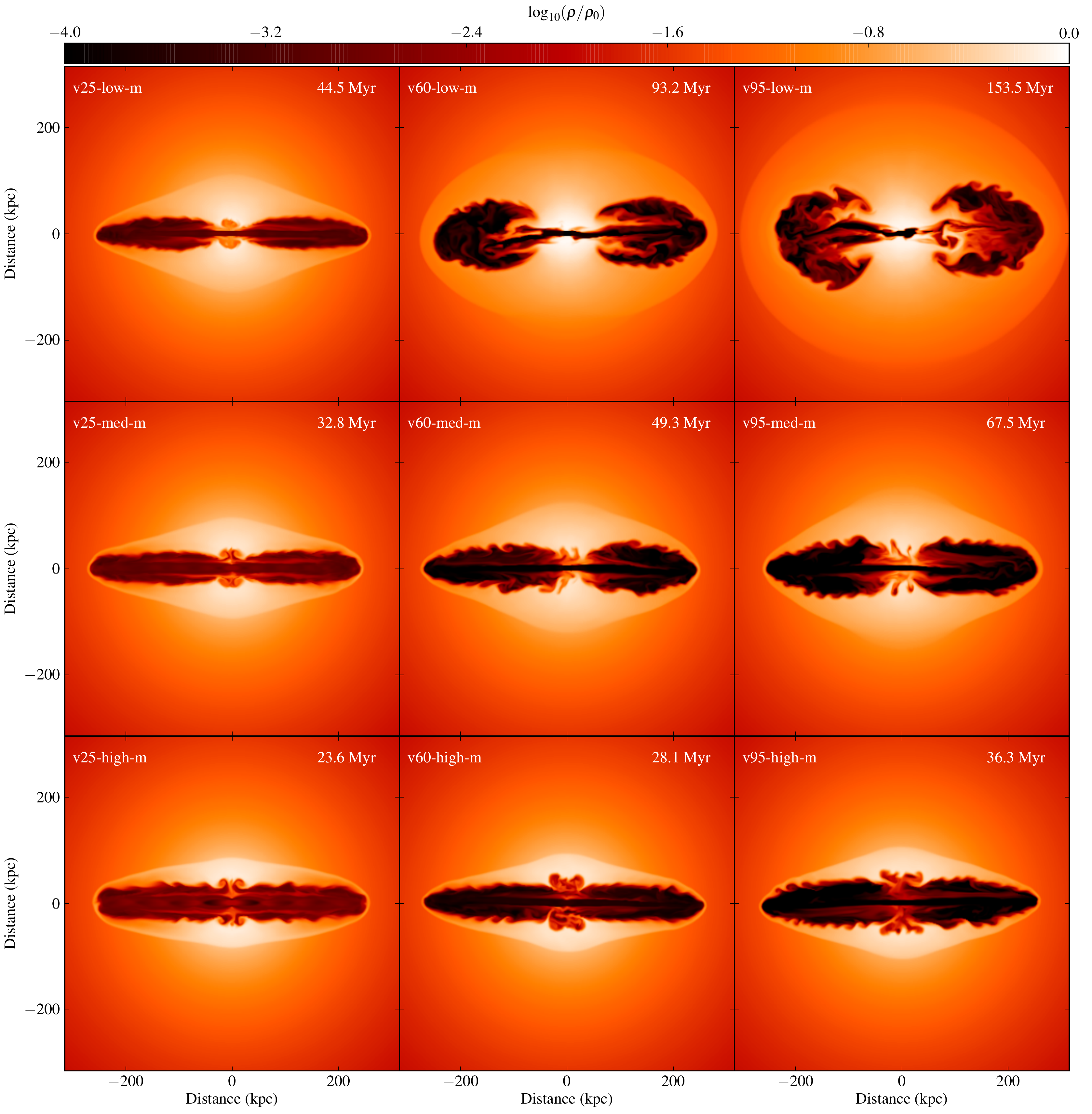}
\caption{Midplane density slices for the suite of RMHD simulations, taken when the average length of the two lobes is 250 kpc. Top row: low power ($Q=1\times10^{39}$ W) models with jet velocities 0.25$c$, 0.6$c$ and 0.95$c$, respectively. Middle row: medium power ($Q=2\times10^{39}$ W) models with jet velocities 0.25$c$, 0.6$c$ and 0.95$c$, respectively. Bottom row: high power ($Q=5\times10^{39}$ W) models with jet velocities 0.25$c$, 0.6$c$ and 0.95$c$, respectively. Colour scale is logarithmic in simulation units of density, ranging from $-4$ (black) to $0$ (white). The label in the top left corner of each plot gives the age of each model at the time of the snapshots.}
\label{fig:RMHDdens}
\end{figure*}

\begin{figure*}
\includegraphics[width=175mm]{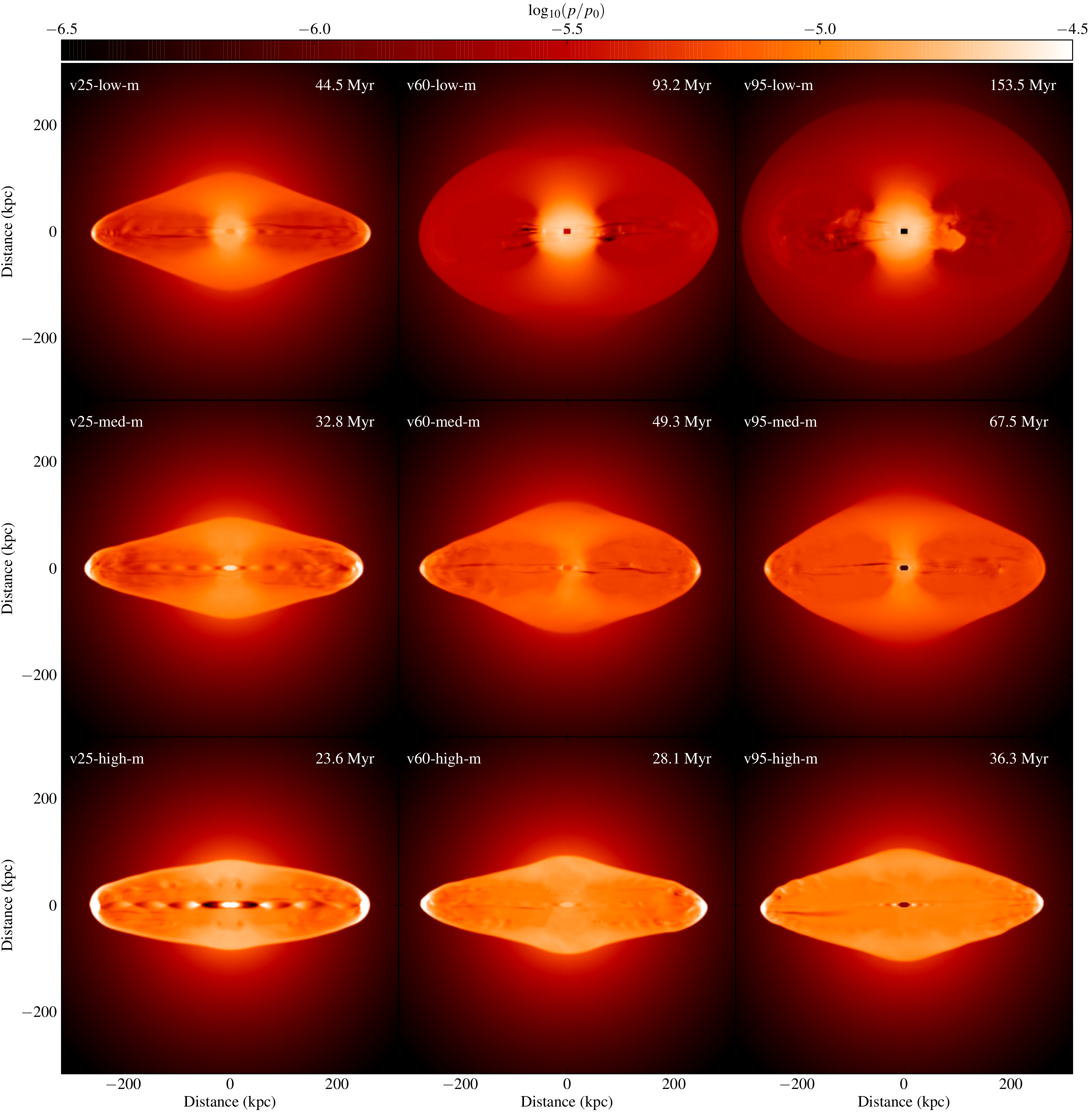}
\caption{Midplane pressure slices for the suite of RMHD simulations, taken at the same time as Fig.~\ref{fig:RMHDdens} and covering the same jet parameters. Colour scale is logarithmic in simulation units of pressure, ranging from $-6.5$ (black) to $-4.5$ (white).}
\label{fig:RMHDprs}
\end{figure*}

\begin{figure*}
\includegraphics[width=175mm]{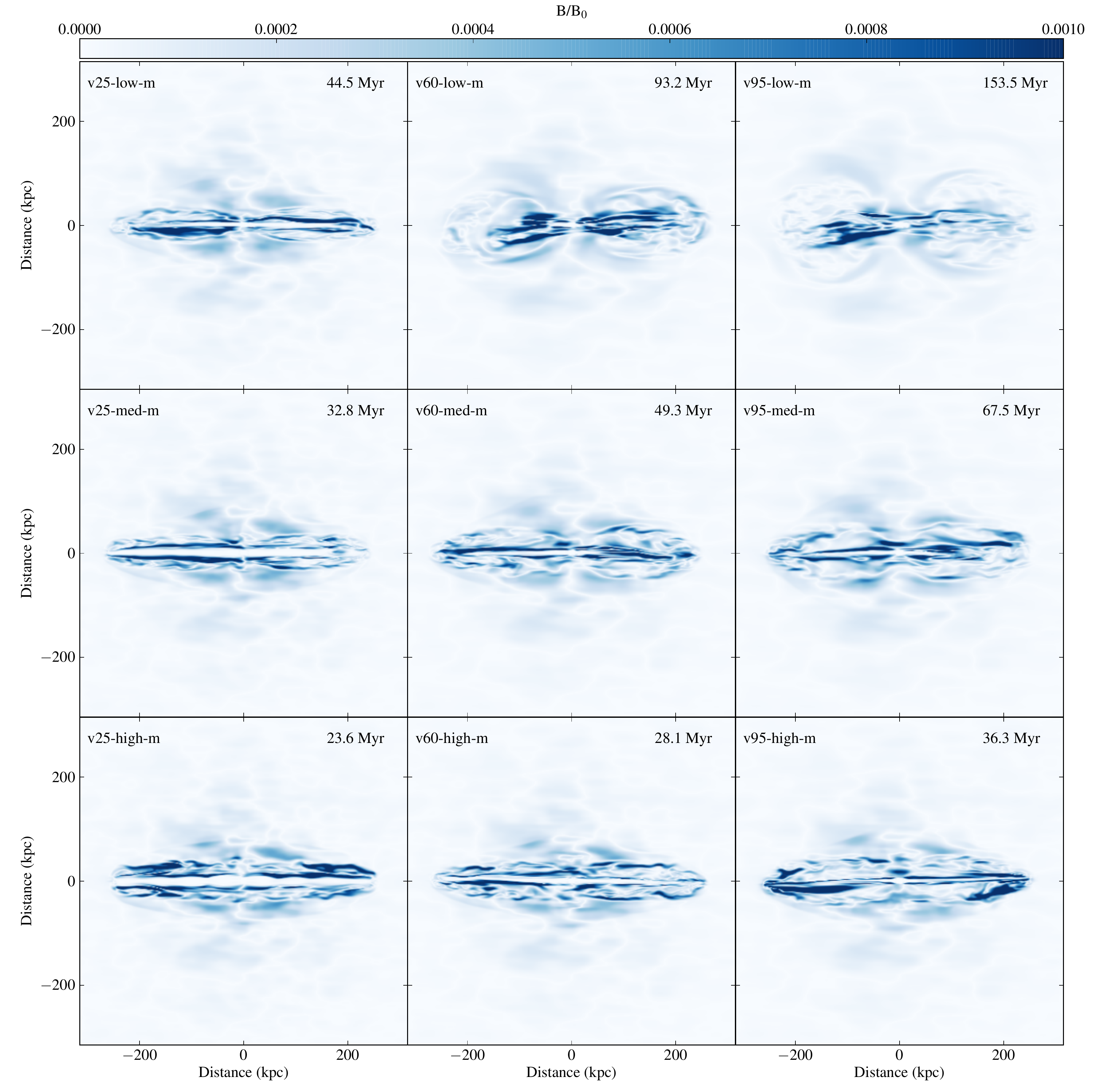}
\caption{Midplane magnetic field strength slices for the suite of RMHD simulations, taken at the same time as Fig.~\ref{fig:RMHDdens} and covering the same jet parameters. Colour scale is in simulation units of magnetic field strength, ranging from $0.0$ (white) to $0.001$ (blue).}
\label{fig:RMHDb}
\end{figure*}

\begin{figure*}
\includegraphics[width=175mm]{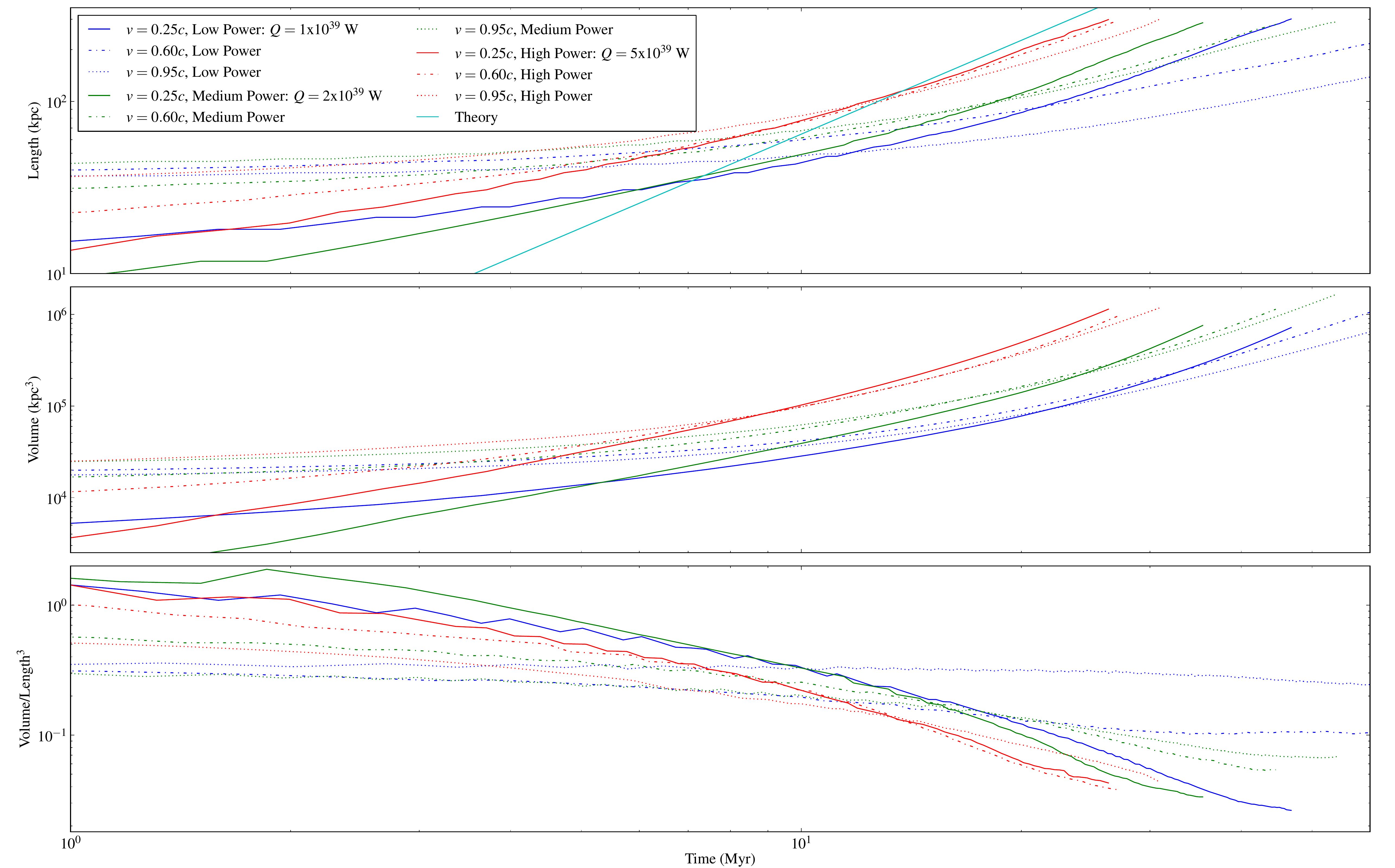}
\caption{Growth of the lobes with time for the RMHD models. Time $t = 0$ is taken to be the time that the model is well coupled with the environment. For each model the plotted value is the average value for the two lobes. The theory line is the predicted growth of the lobe length from \citetalias{KA97}.}
\label{fig:RMHDdyn}
\end{figure*}

Fig.~\ref{fig:RMHDdyn} shows that the lobes follow the same growth as seen for the RHD models with the slow dense jets expanding faster with their higher momentum flux. The length and volume of the lobes initially grow linearly until steepening once leaving the denser core of the cluster, approaching the slope predicted from theory \citepalias{KA97}. The lobes are again seen to be quasi-spherical at early times, as the dense material at the centre of the cluster slows the jets so that the lateral expansion of the lobes is comparable to the longitudinal growth. The presence of the magnetic field has not noticeably damped the Kelvin-Helmholtz instabilities that again appear around the edge of the lobes, presumably because it is not dynamically important. An additional run was performed, with the same initial conditions as the \textbf{v95-med-m} run but with double the injected magnetic energy density. This was also not seen to significantly damp the Kelvin-Helmholtz instabilities.

\begin{figure*}
\includegraphics[width=175mm]{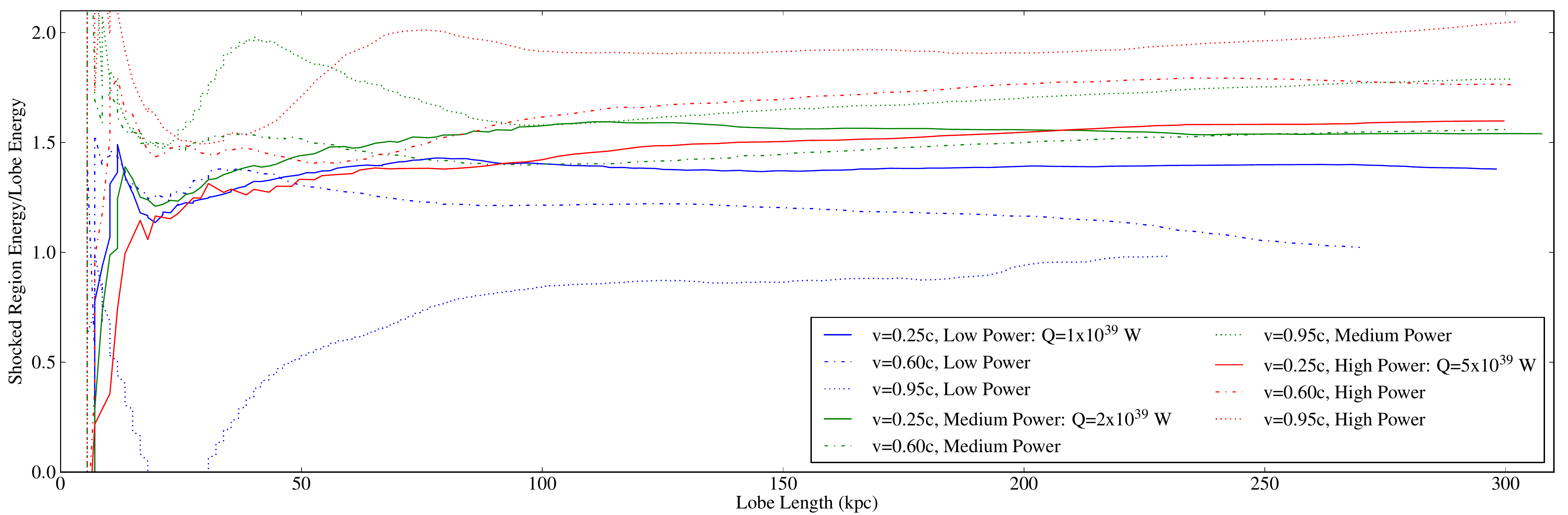}
\caption{Evolution of the ratio of energy stored in the shocked region to that in the lobes, as a function of the length of the lobes, for the RMHD models.}
\label{fig:RMHDenergy}
\end{figure*}

The ratio of energy stored in the lobes to that in the shocked ICM material for the RMHD models, as shown in Fig.~\ref{fig:RMHDenergy}, shows a larger spread than the RHD models, though remains within the 0.6 to 1.8 range seen for the MHD models, except the \textbf{v95-high-m} which is slightly higher. The increased spread in this ratio for the RMHD models can be attributed to the wider range of jet densities used for these models. Since the spread in the MHD models of \citetalias{HK14} is due to the different environments used for the different models it can be expected that, as the lighter jets will be more significantly affected by the environment, this greater spread in the jet densities will lead to a greater spread in the ratio of energy stored in the lobes to that in shocked region.

\begin{figure}
\includegraphics[width=84mm]{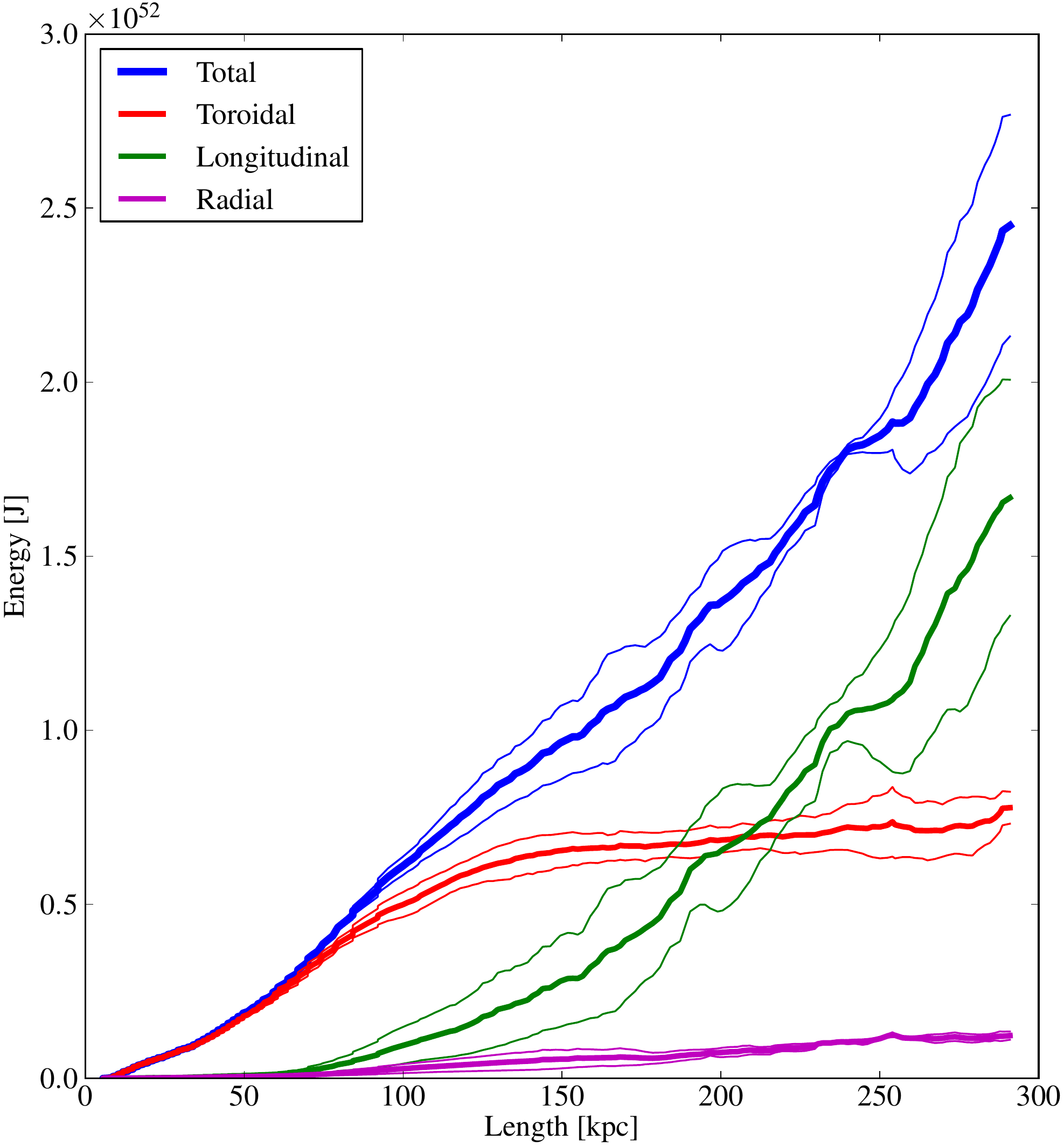}
\caption{Total magnetic energy stored in the lobes of the \textbf{v95-med-m} simulation, and the contribution to this energy from the different components of the magnetic field. The thick lines are energies averaged over the two lobes, and the thin lines are the energies for the lobes individaully to give a sense of the scatter between the two lobes.}
\label{fig:magcont}
\end{figure}

While the magnetic field is injected in a purely toroidal configuration, the momentum of the jet stretches and shears the field along the jet axis leading to the growth of the longitudinal component. At late times the energy stored in the longitudinal component of the magnetic field becomes comparable to, or greater than, the energy in the toroidal component, with the slower models tending to produce a field structure where the longitudinal component is more dominant (Fig.~\ref{fig:magcont}).

\begin{figure*}
\includegraphics[width=175mm]{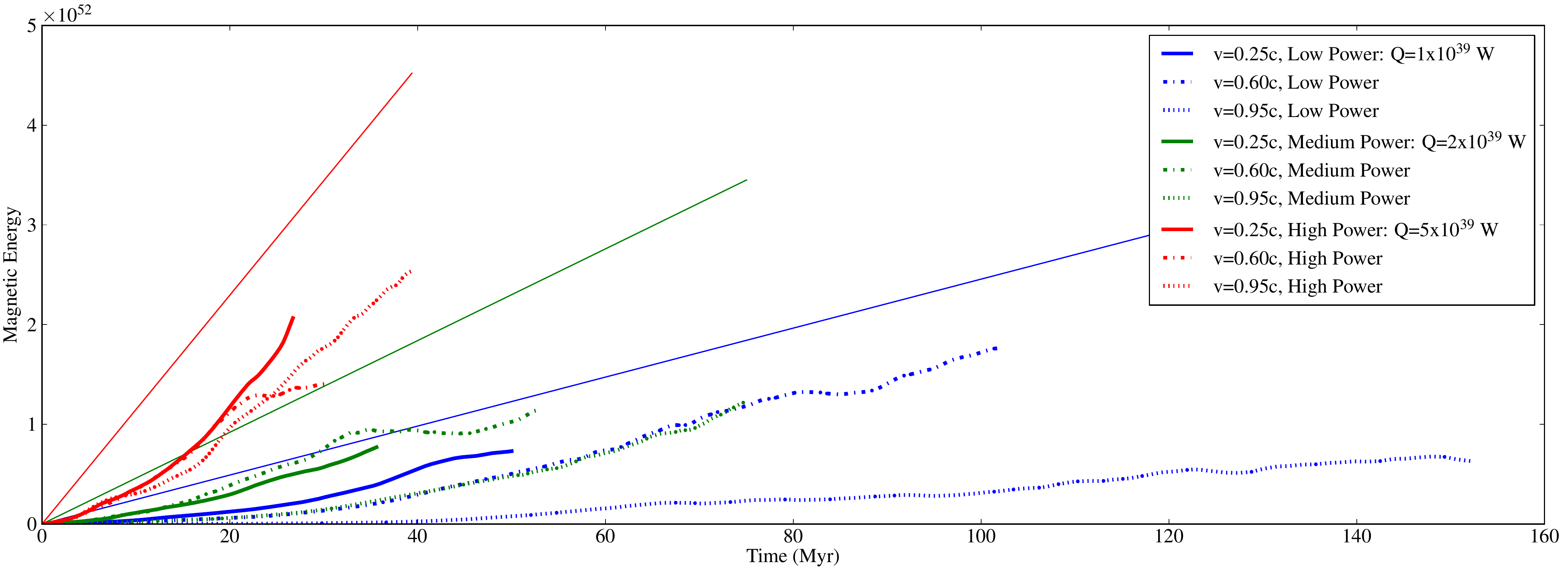}
\caption{Total magnetic energy stored in the lobes of the RMHD models, as a function of time. Thin lines show the predicted amount of magnetic energy in lobes, provided all of the injected magnetic energy makes it onto the grid and that the magnetic fields are not being amplified.}
\label{fig:lobeamp}
\end{figure*}

Fig.~\ref{fig:lobeamp} shows how the amount of magnetic energy accounted for in the simulations differs from the amount injected. As mentioned in section 2, the fact that the energy present is systematically lower than the expected amount is due to the poor coupling between the internal injection region and the ambient medium as well as back-flowing material vanishing into the injection region at early times. It is more useful to compare the \textit{gradient} of the lines. Apart from the \textbf{v95-low-m} and \textbf{v95-med-m} runs, we see that for all of the models the gradient agrees fairly well with the predicted value. For these two runs there is no dominant large-scale field structure in the lobes, so it is possible that there should be field structure on a scale below the resolution of these models resulting in some cancellation of magnetic fields and therefore reduced magnetic energy. We see that at times the gradient of the magnetic energy is greater than the predicted amount which could be initial evidence for magnetic field amplification in lobes of high power jets, though longer simulations of the lobes is required for a conclusive answer. This means that measurements of the magnetic field strengths in the lobes of real radio galaxies could be used to constrain the magnetic field strength around the accretion region. A worry in this context is that unresolved small-scale turbulence could lead to amplification in real sources. This argument is certainly true for the jets, for which our resolution is low. Our simulations do, however, capture the MHD processes in the lobes that re-orient the field structure into a configuration similar to the observed one. Changing the resolution in our non-relativistic simulations (although only by a factor of 1.5) did not change the field amplification significantly. Our resolution is also better than in \citet{HEKA11}, who report similar results. Hence, we believe the small amplification we see in the lobes is realistic, but our results do not constrain amplification in the possibly turbulent jets.

Having seen that the RHD and RMHD models agree in the terms of the evolution of the radio lobes, we can conclude that including weak magnetic fields in the models did not affect the dynamics of the lobes, as expected. This means that we can confidently talk about the RMHD models exclusively for the remainder of this paper. Comparing the results of the RMHD models with the previous models of \citetalias{HK14} we see that running the code with \textsc{pluto}'s relativistic modules does not significantly affect the growth of the lobes, meaning that the results of the previous models are still valid and the effect of environment does not need to be re-investigated in RMHD mode, though further testing with higher Lorentz factor jets are needed to confirm this. While relativistic effects have little impact on the dynamics of the lobes they will affect synthetic observations by boosting the emission from the jet pointing towards the observer and suppressing emission for the jet pointing away.

\subsection{Synchrotron Visualization}

The inclusion of magnetic fields in the models allows us to calculate the Stokes synchrotron emissivities for each cell in the simulation grid. Since the synchrotron emission is anisotropic an angle from which to observe the source must be chosen, in the form of a projection vector pointing from the centre of the simulation volume to the observer. Due to the high velocity of some of the jets in these models the effects of relativistic aberration must be taken into account, where the apparent position of an observer in the reference frame of an object moving at relativistic speeds differs to the position of the observer in the lab frame, by transforming the projection vector into the reference frame of a given simulation cell. The magnetic field components perpendicular to this aberration-corrected projection vector, $B_x$ and $B_y$, are then calculated and used to compute the Stokes $I$ (total intensity), $Q$ and $U$ (polarized intensities) parameters (in simulation units) using the following equations:

\begin{equation} \label{eq:SI}
j_I=p(B_x^2+B_y^2)^{\frac{\alpha-1}{2}}(B_x^2+B_y^2)D^{3+\alpha}
\end{equation}

\begin{equation} \label{eq:SQ}
j_Q=\mu p(B_x^2+B_y^2)^{\frac{\alpha-1}{2}}(B_x^2-B_y^2)D^{3+\alpha}
\end{equation}

\begin{equation} \label{eq:SU}
j_U=\mu p(B_x^2+B_y^2)^{\frac{\alpha-1}{2}}(2B_xB_y)D^{3+\alpha}
\end{equation}

where $p$ is the local thermal pressure, proportional to the number density of electrons for a fixed power-law electron energy distribution, $\alpha$ is the power-law synchrotron spectral index (taken to be $\alpha = 0.5$) and $\mu$ is the maximum fractional polarization (equal to $\mu = 0.69$ for $\alpha = 0.5$). $D$ is the Doppler factor, given by:

\begin{equation} \label{eq:Doppler}
D=\frac{1}{\gamma(1-\beta\cos(\theta))}
\end{equation}

where $\beta = v/c$ and $\theta$ is the angle between the projection vector and the velocity vector of the cell. This is raised to the power $(3+\alpha)$ to account for the increased rate at which photons are received in the lab frame compared to the rate they are emitted, the boosting of these photons to higher energies and the fact that the emitted radiation is preferentially beamed towards the direction of motion. These synchrotron intensities can be converted to physical units by multiplying by the simulation unit of radio luminosity $j_0$, which is given by a modified form of the equation from \citetalias{HK13}:

\begin{equation} \label{eq:SyncPhys}
j_0 = c(q)\frac{e^3}{\epsilon_0 c m_e} \left(\frac{\nu m_e^3 c^4}{e}\right)^{-\frac{q-1}{2}} \frac{3p_0}{4\pi I}\left(\frac{B_0^2}{8\pi\mu_0}\right)^{\frac{q+1}{4}} L_0^3
\end{equation}

where $c(q)$ is a dimensionless constant of the order $\approx 0.05$, $e$ is the charge of an electron and $m_e$ is its mass, $\epsilon_0$ and $\mu_0$ are the permittivity and permeability of free space, respectively, and $c$ is the speed of light. $p_0$, $L_0$ and $B_0$ are simulation units of pressure, length and magnetic field strength, respectively. $q$ is the electron energy power-law index (equal to $2$ for a spectral index $\alpha$ of $0.5$), $\nu$ is the frequency the source is observed at and $I$ is the integral over $EN(E)$ between $E_{min}$ and $E_{max}$, with $E_{min} = 10m_ec^2$ and $E_{max}=10^5m_ec^2$. These values give a simulation unit of radio luminosity to be $j_0 = 3.718\times10^{31}$ W Hz$^{-1}$ sr$^{-1}$.

\begin{figure*}
\includegraphics[width=175mm]{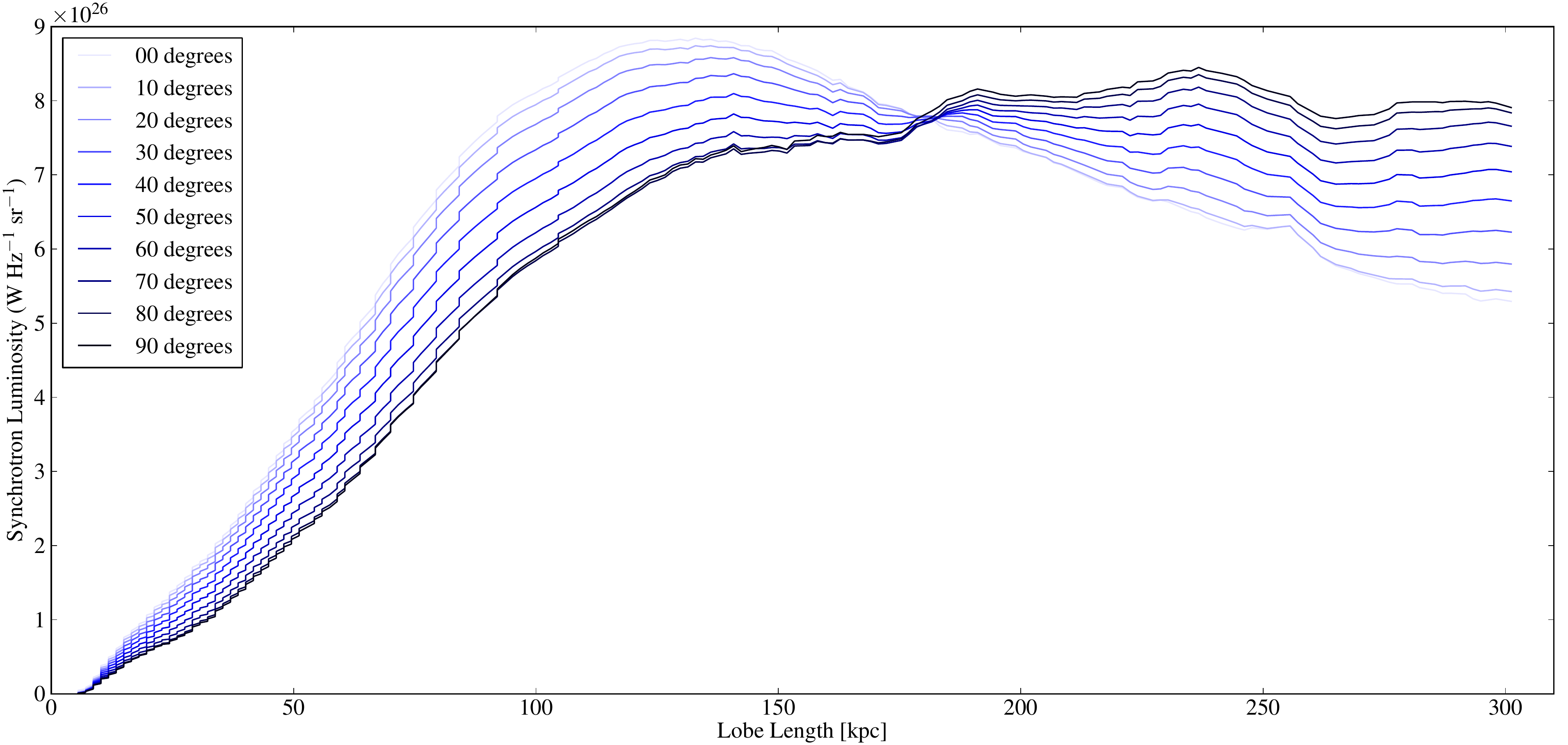}
\caption{Evolution of the synchrotron luminosity with time for the \textbf{v95-med-m} simulation for different viewing angles, where $0$ degrees is parallel to the jet axis and $90$ degrees is perpendicular.}
\label{fig:lightcurve}
\end{figure*}

By integrating this emission over the whole of the source for each output data-cube we can create light-curves for the radio source, and then use different projection vectors to see how the viewing angle affects the observed light-curve. Fig.~\ref{fig:lightcurve} shows this for the \textbf{v60-med-m} simulation, where we see similar evolution to the models of \citetalias{HK14} with the brightness of the source reaching a peak once the length of the lobes reaches around 100-150 kpc. Only the \textbf{v25-high-m} simulation does not follow this track, instead when viewed at angles greater than 30 degrees to the jet axis the brightness rises up until a lobe length of 200 kpc before flattening out. The sources all appear brightest when looking directly down the jet at early times but for all of the models this flips over later so that the source is brightest when looking from the side, with the time at which this flip occurs roughly corresponding to the time that the energy in the longitudinal component of the magnetic field begins to dominate over the energy in the toroidal component. This is likely due to the synchrotron emission being anisotropic, being highest when the magnetic field is perpendicular to the line of sight and with no emission when it is parallel. At early times the magnetic field is mostly toroidal so that when looking along the jet axis the field is entirely perpendicular to the line of sight, whereas when viewed edge on part of the field is parallel and so does not contribute to the luminosity. The longitudinal component has the opposite effect; it does not contribute at all to the emission when observed along the jet axis and is completely perpendicular when looking edge on. For the faster jets Doppler boosting reduces this effect at late times by increasing the brightness when looking down the jet, which has the effect of reducing the scatter between the different lines of sight. At late times we see that the radio luminosity is lower by, on average, $\sim 25$ per cent when the source is viewed along the jet axis when compared to being viewed perpendicular to the jet axis.

\begin{figure*}
\includegraphics[width=175mm]{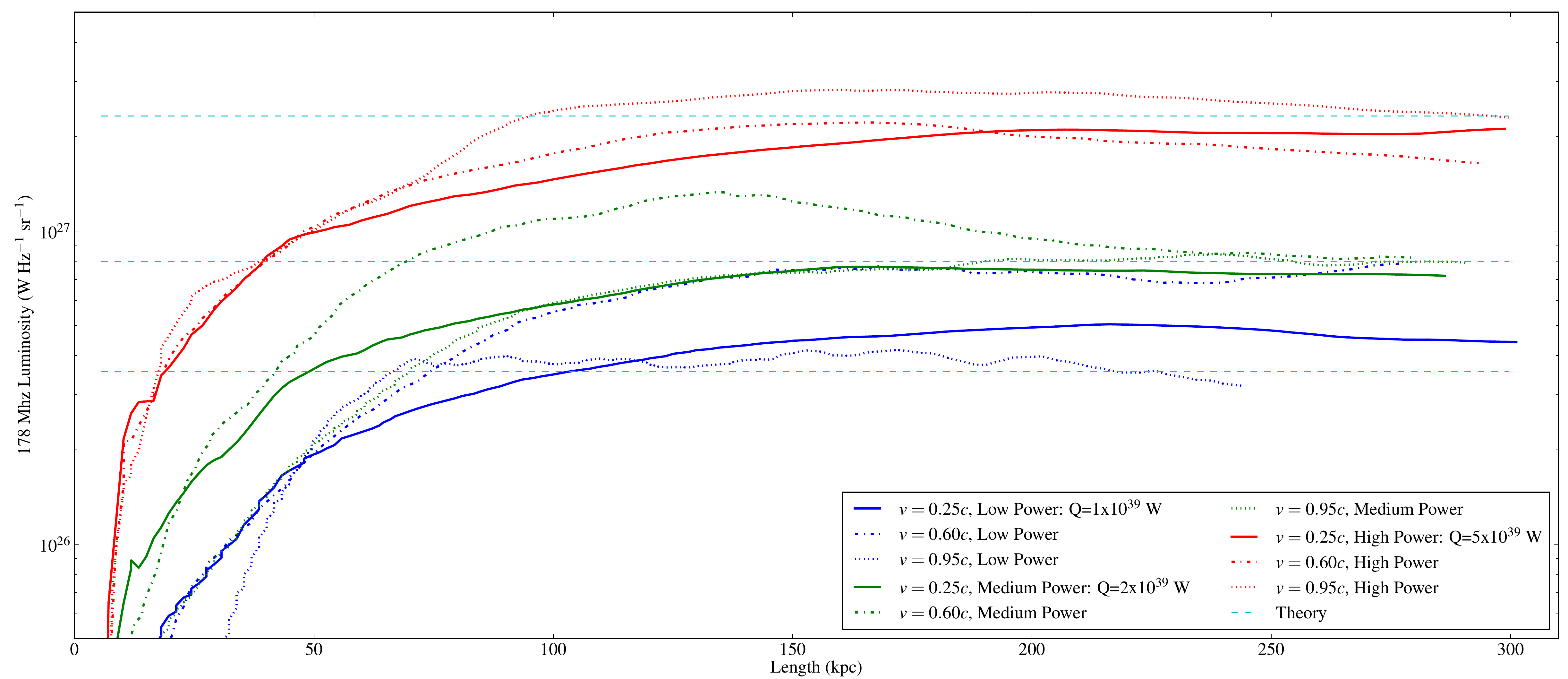}
\caption{Evolution of the synchrotron luminosity with lobe length for each of the RMHD models, for a line of sight $90$ degrees to the jet axis. Luminosities are plotted against lobe length to allow comparison between the models. The theory lines are the radio luminosities predicted for a $1\times10^{38}$ (bottom), $2\times10^{38}$ (middle) and $5\times10^{38}$ (top) W jet using the results of \citet{W99}, with an $f$ factor of 15.}
\label{fig:lightcurves}
\end{figure*}

Fig.~\ref{fig:lightcurves} shows the lightcurves for all of the RMHD models, for a viewing angle of 90 degrees to the jet axis. As expected we see that the higher power jets lead to more luminous sources, with luminosities that are comparable to those expected from the relationship between jet power and 151-MHz radio luminosity of \citet{W99}, with the best agreement being for an $f$ factor (a factor included to account for systematic uncertainties, which has a value greater than unity) of 15, within the suggested range (10 to 20) of \citet{BR00}. While there is a small amount of scatter between models of equivalent jet power, no relationship between jet speed and radio luminosity is seen.

\begin{figure*}
\includegraphics[width=175mm]{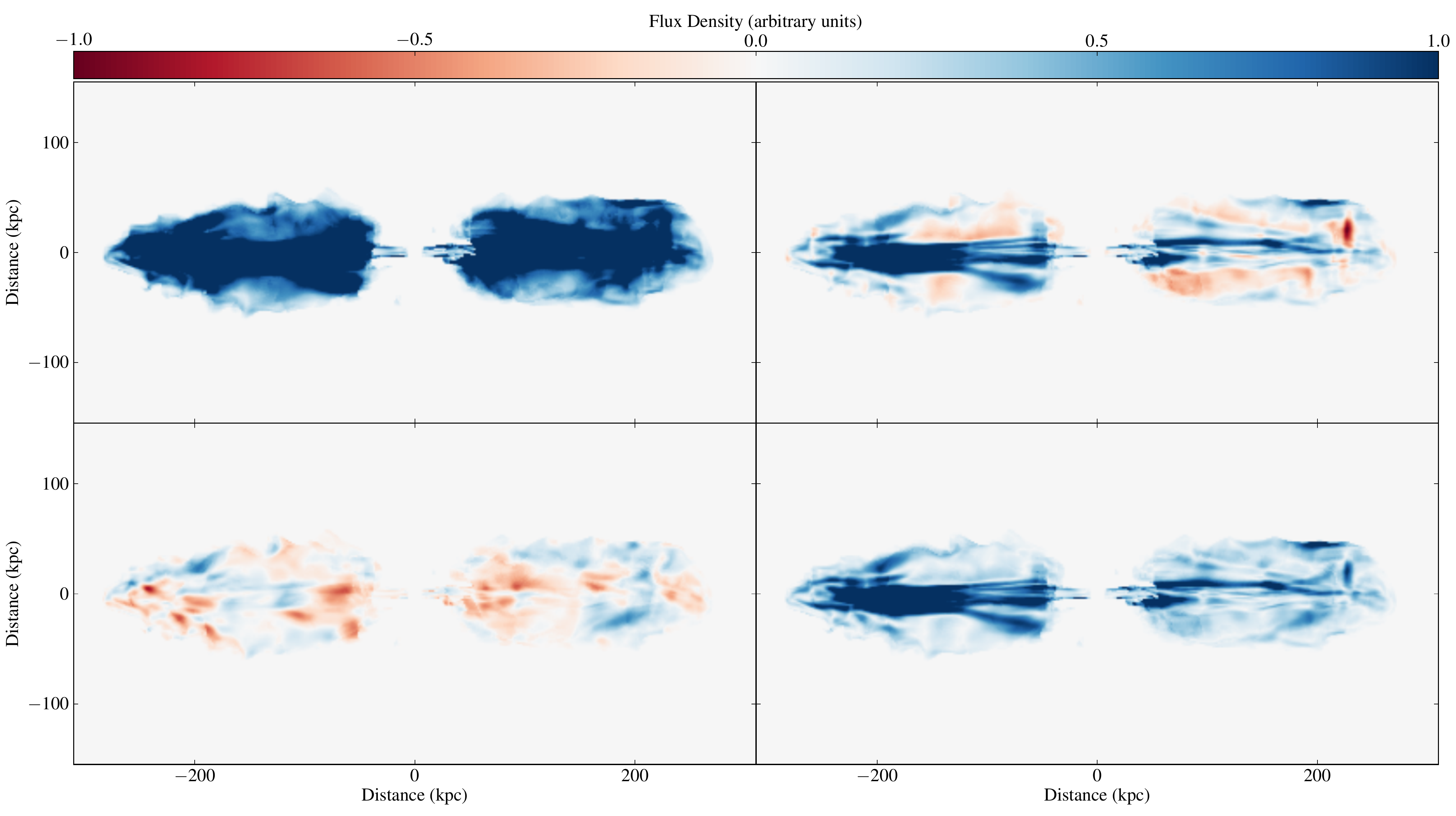}
\caption{Synchrotron emission maps for the \textbf{v95-med-m} simulation, observed at 90 degrees to the jet axis at an age of 67.5 Myr. Top row: Stokes $I$ (left) and $Q$ (right). Bottom row: Stokes $U$ (left) and $P=\sqrt{Q^2+U^2}$ (right). All maps are scaled by the same arbitrary amount so that faint structure can be seen in all of the maps.}
\label{fig:synchmaps}
\end{figure*}

Instead of integrating over the whole source we can integrate along lines of sight to create two-dimensional emission maps, show in Fig.~\ref{fig:synchmaps}. We see in all of the maps regions of very strong emission alongside the jet resulting from the strong magnetic fields being sheared by the jet, similar to the structure seen in \citetalias{HK14} though stronger here due to the higher jet speeds. The polarized intensities also show the patchy emission previously seen, attributed to a highly complex magnetic field structure once the simulations have been allowed to evolve to late times. For the models which have persistent structure between the lobes from early back flowing jet material, emission is seen to be very high from this region. The hotspots seen in observed radio lobes resulting from the jet ending in a termination shock are only seen in the slower, higher power models at early times, and at late times only the \textbf{v25-high-m} simulation still has visible hot spots.

\begin{figure*}
\includegraphics[width=175mm]{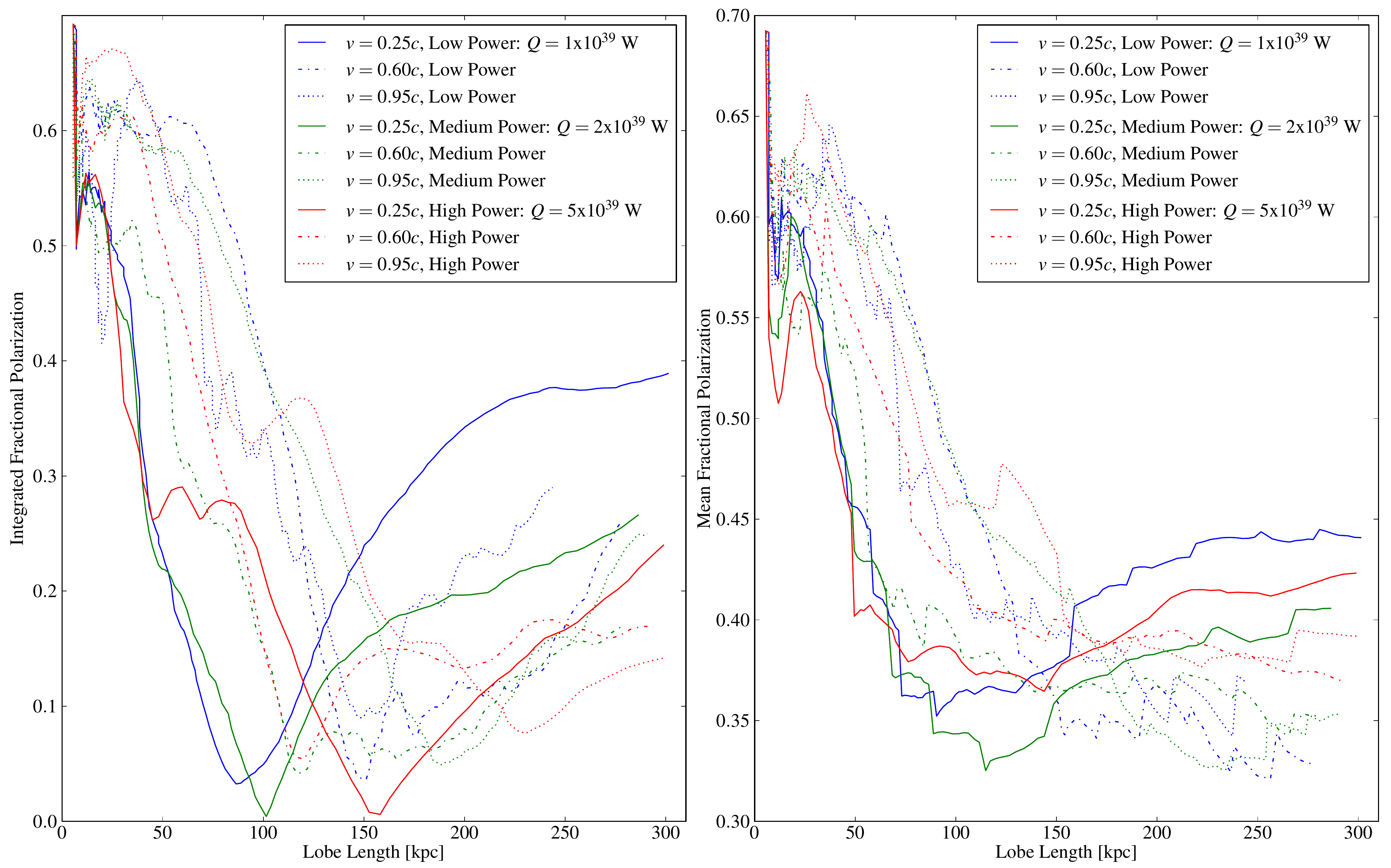}
\caption{Integrated (left) and mean (right) fractional polarizations as a function of lobe length for the RMHD models, viewed at an angle 90 degrees to the jet axis. Note the different $y$ axis scaling used between these two plots.}
\label{fig:polar}
\end{figure*}

We can also look at the fractional polarization  (the fraction of the synchrotron emission that is in the polarized $Q$ and $U$ intensities) and how this fraction evolves for the different models, neglecting the effects of Faraday rotation. Fig.~\ref{fig:polar} shows two forms of the fractional polarization. The first is the integrated fractional polarization (left panel of Fig.~\ref{fig:polar}), $F_{tot}=\sqrt{Q_{tot}^2+U_{tot}^2}/I_{tot}$, which is what would be measured for an unresolved source, where $I_{tot}$, $Q_{tot}$ and $U_{tot}$ are the Stokes parameters integrated over the whole source (where $I$ is not zero). We see that all of the models follow the same trend of initially decreasing down to a minimum value, which occurs at roughly the lobe length at which the magnetic field structure switches from being strongly toroidal to predominantly longitudinal. During the initial decline we see that the slower jets have a significantly lower fractional polarization than the faster jets, due to the strong shearing of the magnetic field in these models.

The right-hand panel of Fig.~\ref{fig:polar} shows the mean fractional polarization, where $F$ is calculated in a similar way to above, but on a pixel-by-pixel basis at the full numerical resolution of the models, as would be measured for a well resolved source. The trend for all of the models is the same as for the unresolved source, although with higher values at all times and with less pronounced minima, with an overall decrease with time as the magnetic field becomes increasingly disordered. Again the slower jets are seen to have higher fractional polarizations at late times, as the stronger shear now produces a more dominant longitudinal field structure. The values seen here for both mean and integrated fractional polarization are very similar to values measured for the MHD models of \citetalias{HK14} and \citet{HEKA11}.

\section{Discussion}

Here we compare our models to a sample of observed radio galaxies, and to an analytic model describing the relationship between lobe and jet radius, in order to assess how well these results can be believed.

\begin{figure*}
\includegraphics[width=175mm]{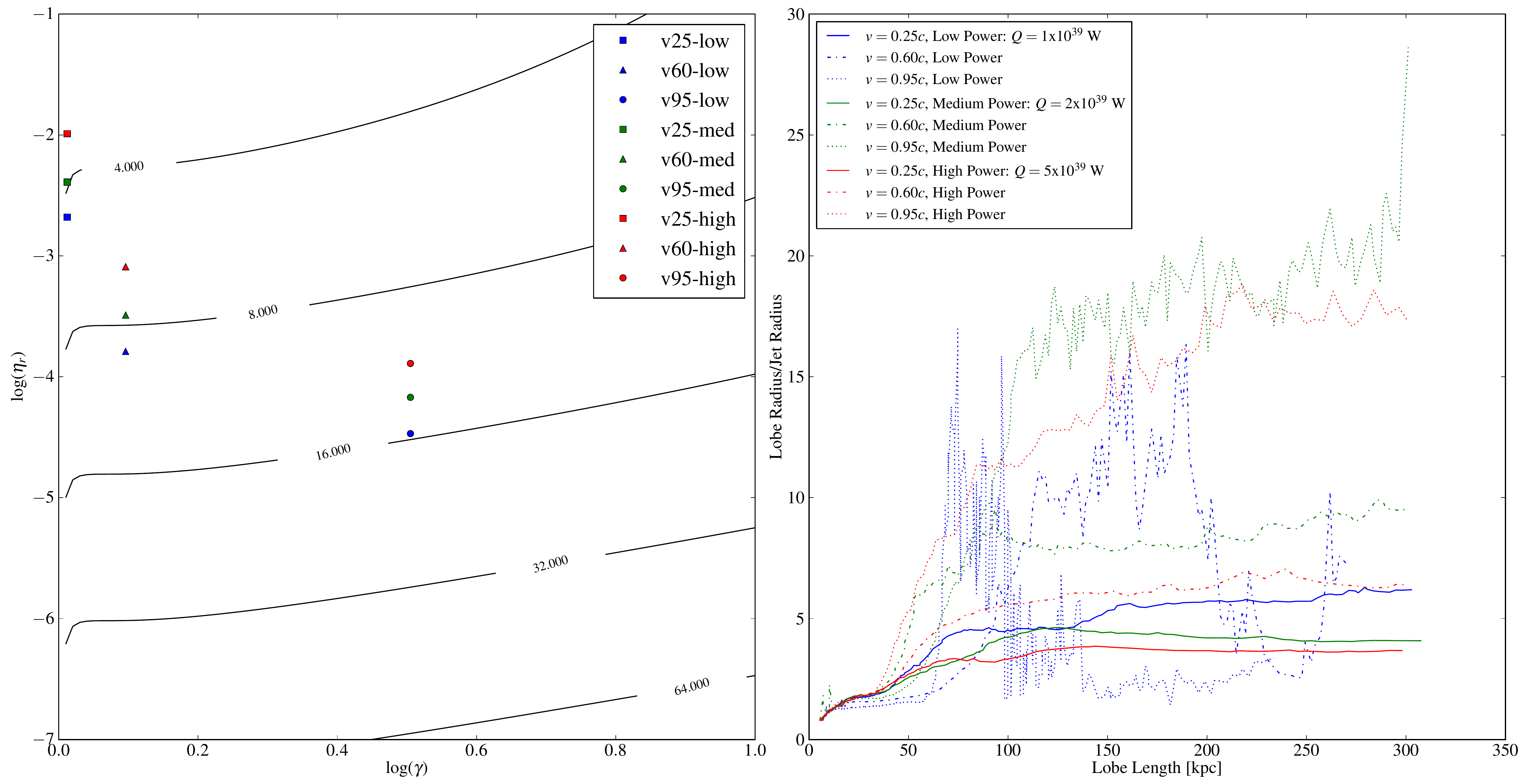}
\caption{Left: Predicted values for the ratio of lobe radius ($R_l$) to jet radius ($R_j$), as a function of $\gamma$ and $\eta_r$. Contours of equal $R_l/R_j$ are calculated from the analytical model of \citet{R99}. The points correspond to each of the RMHD models, using injected parameters and conditions at the centre of the cluster to calculate $\eta_r$. Right: Average measured value of $R_l/R_j$ for the RMHD models, as a function of lobe length.}
\label{fig:radratio}
\end{figure*}

Fig.~\ref{fig:radratio} shows a comparison between our models and the simple analytical model presented by \citet{R99}. This model predicts a value for the ratio of lobe radius ($R_l$) to jet radius ($R_j$) based upon the jet's values of $\gamma$ and $\eta_r$. Contours show lines of constant $R_l/R_j$, and the points show the predicted ratio for each model based upon the injected jet properties and the value of $\eta_r$ calculated at the centre of the cluster. As the atmosphere in our models is not uniform, the predicted value of $R_l/R_j$ is also non-uniform and decreases slowly with radius, meaning that the calculated value for this ratio is expected to be slightly lower than the predicted value. We see that most of our models agree well with this. The \textbf{v95-med-m} and \textbf{v95-high-m} have higher values than predicted, and the \textbf{v95-low-m} run shows spurious values that are much lower than expected. This is due to the jet in these models being disrupted and dissipating very early on in the lobes, resulting in the tracer value in the jet falling below the threshold to be identified as jet material as opposed to lobe material. For the first two cases this results in the jet radius being calculated to be much lower than the radius of the injection region, giving a higher value for $R_l/R_j$ for the same lobe radius. For the \textbf{v95-low-m} model the disruption of the jet is much more significant, such that only the first few kpc of the jet are identified. At late times when the lobes have been pushed away from the centre no values for the ratio are calculated in the lobes themselves, which results in a ratio of $\sim$$1$ at late times.

\begin{figure*}
\includegraphics[width=130mm]{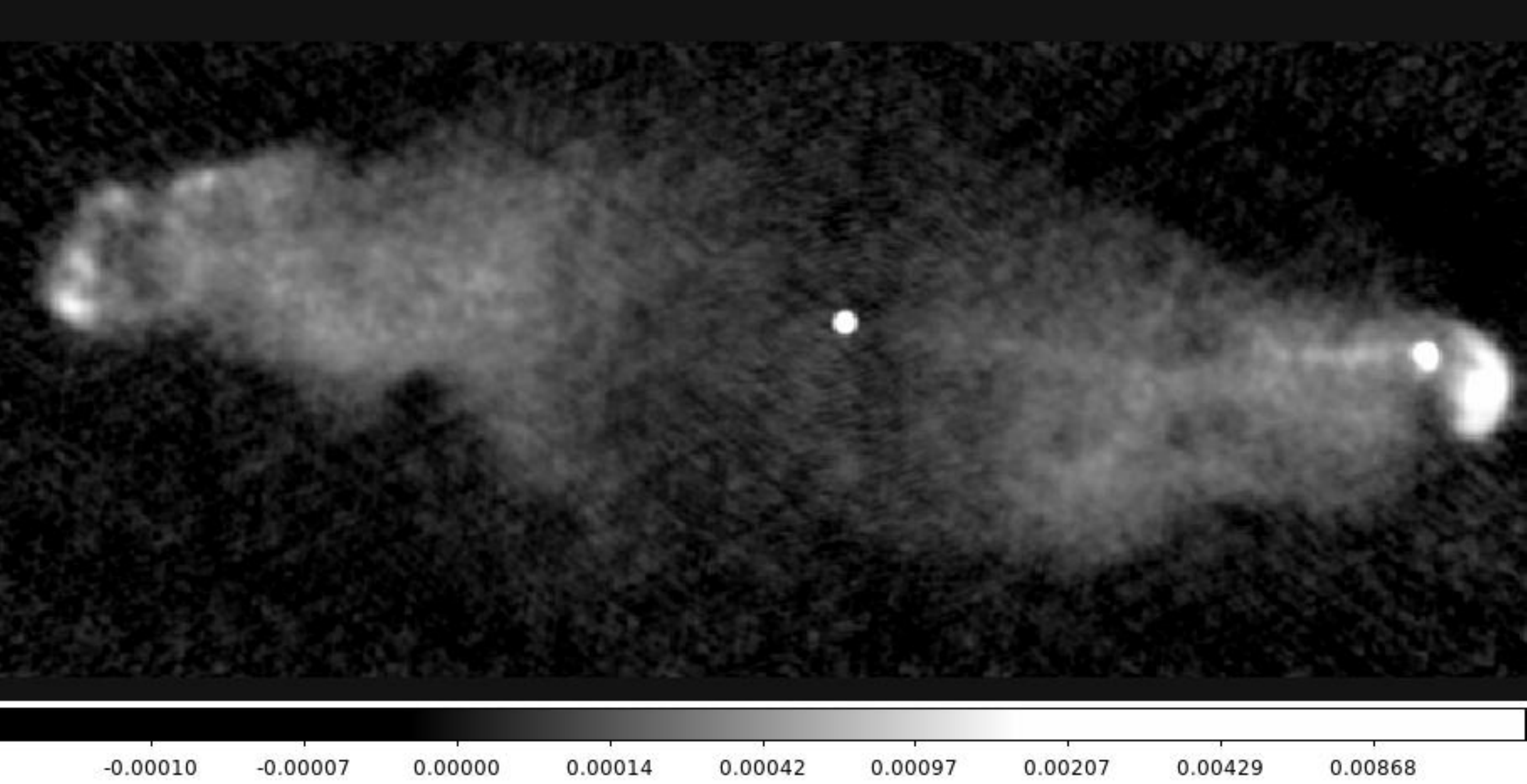}
\caption{Total-intensity map of radio galaxy 3C436, from \citet{H97}, with 0.75 acrsec resolution and a logarithmic colour scale.}
\label{fig:3C436}
\end{figure*}

Fig.~\ref{fig:3C436} shows a total-intensity radio map for the typical radio galaxy 3C436 from \citet{H97}. Comparing this to the synthetic observation for the \textbf{v95-med-m} run (Fig.~\ref{fig:synchmaps}) we see that many of the same features are present; lobes that are symmetric on the large scale and are expanding away from the central source, resulting in the emission being 'pinched in' around the centre, as well as the emission from the jets on one side of the source appearing as a broken line connecting the central source to the edge of the lobes. The main feature missing from our synthetic observations is the bright hotspots at the end of the lobes, which are observed for almost all radio galaxies, but only seen in the \textbf{v25-high-m} run.

\begin{figure*}
\includegraphics[width=175mm]{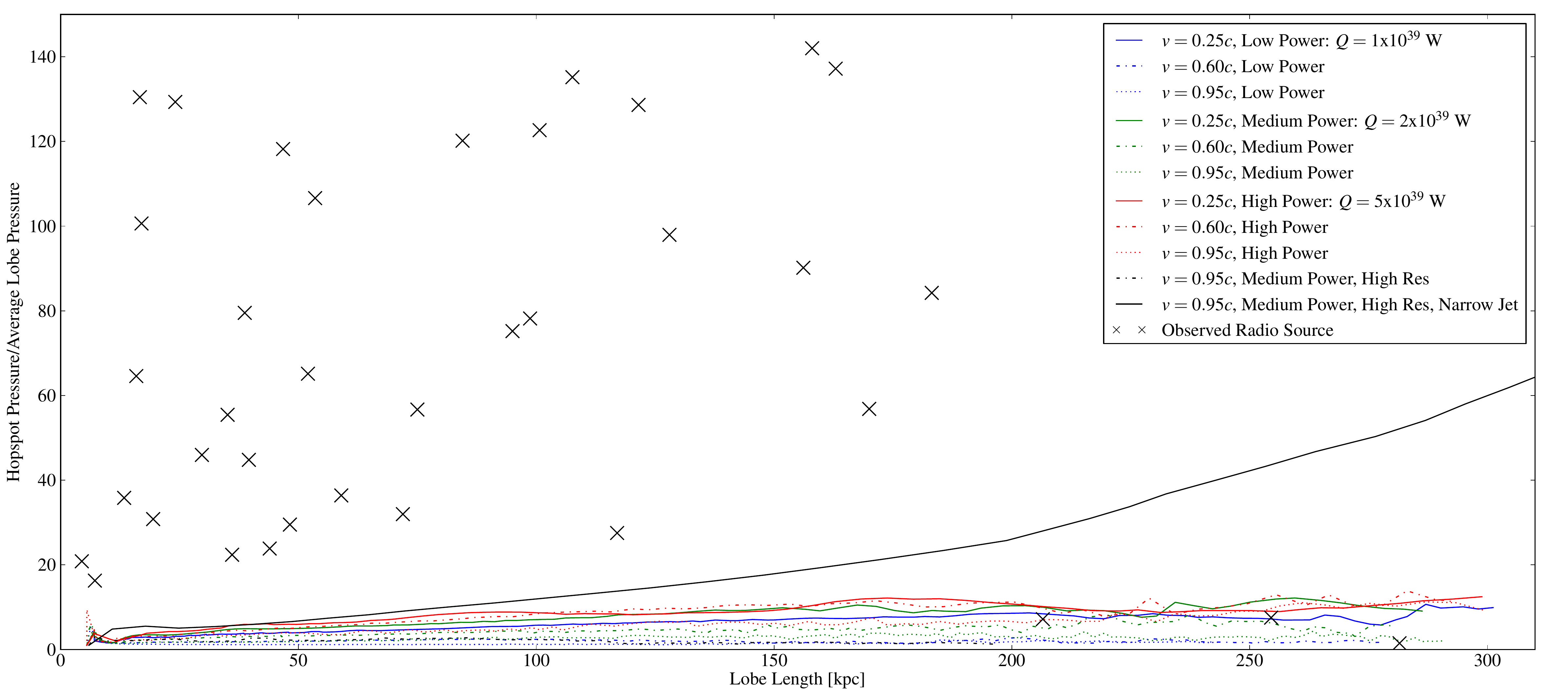}
\caption{Ratio of the pressure at the hotspots to the average pressure in the lobes for observed radio galaxies and for the RMHD models, as a function of lobe length.}
\label{fig:prsrat}
\end{figure*}

In order to investigate the absence of hotspots from the synthetic observations for most of the models the ratio of pressure at the hotspot of the lobes to the average pressure in the lobes was calculated for the RMHD models and for observed radio galaxies (Fig.~\ref{fig:prsrat}). The observations are taken from the \citet{MRH08} sample of radio galaxies with $z < 1$, and pressures are calculated from the radio luminosity and estimated volume (assuming spherical hotspots and ellipsoidal lobes) of the different regions and by assuming equipartition between the energy stored in the particles and that in the magnetic fields. For the simulations the hotspot pressure is taken from the cell with the highest thermal pressure. We see that all of our models have much lower pressure ratios than the majority of the observed radio sources. To test whether this was the reason for the missing hotspots two further simulations were run, at double the resolution of the rest of the models (achieved by simulating a smaller volume such that only one of the lobes is modelled). The first of these models uses the same injection parameters as the \textbf{v95-med-m} run. The dynamics, energetics and calculated pressure ratio of this run are seen to be very similar to the low resolution run, suggesting that the lack of hotspots is not due to the termination shock not being resolved. The second uses the same jet power and velocity but with the jet radius reduced by a factor of 2 with the injected energy density adjusted accordingly to produce a higher pressure jet that is more comparable, though still too wide, in terms of jet radius to observed sources. The pressure ratios seen for this model are much more comparable to the observed sample, but we still do not see the hotspots in the synthetic observations. We conclude from this that we only observe hotspots in the synthetic observations of models that have properties that are not seen in real radio sources, such as very wide ($>10$ kpc) jets or jets that are over-pressured with respect to the cluster core. In order to recreate observed hotspots in these models we must therefore include additional physics in our models such as particle acceleration in shocks. Another path to explore would be to extend the parameter space further by running models with higher magnetic field strengths, Mach number jets and resolutions.

\section{Summary and Conclusions}

We have performed 3D RHD and RMHD numerical simulations of the evolution of the lobes of radio galaxies in a realistic cluster environment, covering a range of jet powers and velocities, in order to see how the dynamics and emission properties of the lobes depend on the velocity and power of the jets, and to see how the results of Papers 1 and 2 hold up for relativistic jets.

We have seen that for a given jet power the lobes of faster jets expand much slower, since they are necessarily lighter and therefore have a lower momentum flux for the same kinetic energy flux. The result of this is that the faster, lighter jets will inflate significantly wider lobes, staying almost spherical for nearly the whole evolution of the lobes. Other dynamic properties are seen to have little dependence on the lobe advance speed in terms of the overall trend; for all of the models the lobes begin with slow growth as the jet propagate through the dense cluster core, but begin to speed up and approach the expected speed predicted by \citetalias{KA97} as the cluster density falls. The slower, denser jets are seen to have faster lobe expansion at all times. We see reasonable agreement between our models and the analytical model of \citet{R99}. The ratio of energy stored in the lobes to that put into the cluster is seen to be fairly constant regardless of jet power, jet velocity or numerical prescription used. This suggests that we have a robust description of the work done on the cluster by this type of radio source.

Our synthetic synchrotron emissivities are seen to produce values that are in very good agreement with the relationship of \citet{W99} at late times, with the luminosity of all of the sources being flat once the lobes have left the core of the cluster with little spread between different models of the same jet power, though the flat part is a shorter part of the lobes' evolution than is implied by Fig.~\ref{fig:lightcurves} since the growth of the radio lobes is much slower in the central 100 kpc of the cluster. It is worth noting that the synchrotron emissivities presented here do not take into account light travel time (we see the emission from all parts of the source instantly for each output file) or the effects of spectral aging, which would make these light-curves significantly less flat. Doppler boosting is seen to have little effect on the luminosity of these models, since little emission comes from the jets themselves and the emitting material is not moving at highly relativistic speeds. Instead the dependence of luminosity on the viewing angle is due to the structure of the magnetic field. At early times the field is purely toroidal and the source appears brightest when viewed along the jet axis, but as the jet shears the field and the longitudinal component begins to dominate the source appears brightest when viewed perpendicular to the jet axis. At late times the difference in radio luminosity between the different viewing angles is significant, with the source appearing dimmer by $\sim 25$ per cent when looking directly down the jet as opposed to being viewed edge-on. Observed radio lobes are typically seen to have jet-aligned magnetic field vectors. Our results therefore suggest a bias in flux-limited samples towards high inclinations. Calculations based on radio luminosity, such as estimations of jet power, will be incorrect unless this dependence on viewing angle is taken into account.

The polarization properties of the emission are seen to be largely independent of jet velocity, with all of the models following roughly the same evolution of fractional polarization with time. Emission maps of the Stokes parameters are seen to be very similar to those of \citetalias{HK14}, with a filamentary structure seen alongside the jet, especially in the Stokes $P$ maps, and a patchy structure seen in the polarized Stokes $Q$ and $U$ maps which are evidence for a complex magnetic field structure at late times. While some small amplification of the magnetic field is seen (up to a factor of $\sim 2$ in powerful sources), overall the amount of magnetic energy present agrees reasonably well with the injected amount. As with \citetalias{HK14}, we do not see the hotspots in the synchrotron emission resulting from the jet termination shocks for the majority of our models, and conclude that in order to reproduce observed hotspots from models with realistic input parameters we must include additional physics in the form of particle acceleration at shocks. Models with higher Lorentz factor jets could also help, since they will have a higher Mach number, will be more stable and will provide a more consistent supply of energy to the end of the lobes. While we have seen that the velocity of the jet material significantly affects the shape of the lobes, the growth of the lobes follows the same general trend for all of the models in this and previous papers. The emission properties are also seen to be mostly independent of the type of model we run, confirming the results of the previous papers even up to the relativistic velocities used in these models.

Our future work will look to further improve upon these models in order to produce an accurate description of the relationship between observed properties and intrinsic parameters in powerful radio galaxies. Running the models at a higher resolution would allow us to model lower power jets by reducing the size of the injection region, and consequently the width of the jets, to a size more comparable to observed sources. Including the transport and shock acceleration of cosmic rays, radiative losses, and spectral aging effects would all work to create more realistic synthetic observations, and allow better comparison with observations. More realistic cluster environments, as opposed to the spherically symmetric model currently used, could also be implemented by extracting environments from cosmological simulations.

\section*{Acknowledgments}

WE and MJH acknowledge support from the UK's Science and Technology Facilities Council [grant numbers ST/M503514/1 and ST/M001008/1]. This work has made use of the University of Hertfordshire Science and Technology Research Institute high-performance computing facility. This work was supported by funding from Deutsche Forschungsgemeinschaft under DFG project number PR 569/10-1 in the context of the Priority Program 1573 ``Physics of the Interstellar Medium''. We also thank an anonymous referee for useful and constructive comments that helped improve this paper.






\bsp	
\label{lastpage}
\end{document}